\documentclass[prd,twocolumn,nofootinbib,showpacs,superscriptaddress]{revtex4}

\usepackage{amsfonts}
\usepackage{amsmath}
\usepackage{amssymb}
\usepackage{bm}
\usepackage{dcolumn}
\usepackage{epsfig}
\usepackage{graphicx}
\usepackage[latin1]{inputenc}
\usepackage{latexsym}
\usepackage{rotating}
\usepackage{hyperref}
\usepackage[usenames]{color}
\usepackage{float}
\usepackage{xspace} 
\usepackage{mathrsfs}
\usepackage{subfigure}
\usepackage{enumitem}
\usepackage{caption}
\usepackage{ulem}
\definecolor{darkgreen}{rgb}{0.2,0.7,0.2}

\newcommand\be{\begin{equation}}
\newcommand\ba{\begin{eqnarray}}
\newcommand\ee{\end{equation}}
\newcommand\ea{\end{eqnarray}}

\newcommand{\dint}{\displaystyle\int}

\usepackage{setspace}

\begin{document}
\title{Constraining the Solution to the Last Parsec Problem with Pulsar Timing}

\author{Laura Sampson}
\affiliation{Department of Physics, Montana State University, Bozeman, MT 59717, USA.}

\author{Neil J. Cornish}
\affiliation{Department of Physics, Montana State University, Bozeman, MT 59717, USA.}

\author{Sean T.~McWilliams}
\affiliation{Department of Physics and Astronomy, West Virginia University, Morgantown, WV 26506, USA.}

\date{\today}

\begin{abstract} 
The detection of a stochastic gravitational-wave signal from the superposition of many inspiraling supermassive black holes with pulsar timing arrays (PTAs) is likely to occur within the next decade. With this detection will come the opportunity to learn about the processes that drive black-hole-binary systems toward merger through their effects on the gravitational-wave spectrum. We use Bayesian methods to investigate the extent to which effects other than gravitational-wave emission can be distinguished using PTA observations. We show that, even in the absence of a detection, it is possible to place interesting constraints on these dynamical effects for conservative predictions of the population of tightly bound supermassive black-hole binaries. For instance, if we assume a relatively weak signal consistent with a low number of bound binaries and a low black-hole-mass to galaxy-mass correlation, we still find that a non-detection by a simulated array, with a sensitivity that should be reached in practice within a few years, disfavors gravitational-wave-dominated evolution with an odds ratio of $\sim$30:1.  Such a finding would suggest either that all existing astrophysical models for the population of tightly bound binaries are overly optimistic, or else that some dynamical effect other than gravitational-wave emission is actually dominating binary evolution even at the relatively high frequencies/small orbital separations probed by PTAs. 
\end{abstract}
\maketitle
\allowdisplaybreaks[4]

\section{Introduction}
The history of structure formation in our universe is understood to be one of hierarchical building, in which small galaxies merge to form larger galaxies, and large-scale structure emerges in the course of these interactions. It is commonly assumed that this history of galactic mergers implies a companion history of black-hole mergers~\cite{Kormendy:2013dxa}, which suggests the existence of a large population of supermassive-black-hole (SMBH) binaries that emit gravitational radiation. There is some observational evidence that SMBH binaries do, in fact, exist \cite{2011arXiv1106.2952E, 2011ApJ...738...20T}. If the SMBHs are at separations where they emit appreciable energy in gravitational waves (GWs), these GWs will add together to form a stochastic background of gravitational radiation, which can be searched for using pulsar timing arrays (PTAs) \cite{2013CQGra..30v4008M,2010CQGra..27h4014F,2013PASA...30...17M,doi:1955503}.

To search for GWs, PTAs take advantage of the fact that millisecond pulsars are the best natural clocks in the universe. These rapidly spinning neutron stars emit radio waves that sweep past the Earth with stunning regularity - for the best systems, we can predict the time of arrival (TOA) of the radio pulses with accuracies of tens of nanoseconds. These predictions are made using timing models that include the orbits of the pulsars and the Earth, spin-down of the individual pulsars, changes in the instruments, and many other effects~\cite{lorimer2005handbook,Hobbs21122012,2010ApJ...720L.201C}. Once the parameters of this timing model are fitted, the model is subtracted from the TOA data, leaving behind timing residuals. If there is a stochastic background of GW radiation permeating the space between the Earth and the observed pulsars, these timing residuals will include the GW signal, manifesting as a red ``noise'' source. 

In an individual pulsar, the effect of this stochastic background would be impossible to detect, as it could be modeled away as one of the other known red noise processes. However, Hellings and Downs~\cite{HellingsDowns} showed that the stochastic GW signal due to an ensemble of binaries will be correlated between pairs of pulsars, with a correlation function that depends on the relative separation of the pulsars on the sky. It is by searching for red power content that possesses this correlation that PTAs can detect a stochastic GW signal~\cite{2014ApJ...794..141A,doi:1900449,2011MNRAS.414.3117V,2013CQGra..30v4015S}.

It is certainly true that at some point in the inspiral of SMBH binaries, the emission of GWs and the concordant loss of energy and angular momentum becomes the primary driver toward merger. In order for these systems to merge within a time period shorter than the current age of the universe, however, it is not possible for GWs to be solely responsible for the evolution of the SMBH system after the initial galactic merger. There must be other mechanisms that drive the black holes near enough to each other for GW emission to dominate. These mechanisms include phenomena such as stellar scattering and the accretion of gas, which are thought to become less efficient as the black holes approach merger~\cite{2012AdAst2012E...3D,Sesana:2006xw,2005ApJ...634..921A,Ivanov:1998qk,2011ApJ...732...89K,Quinlan:1996vp,2014arXiv1406.5297R}. For the elliptical galaxies that host BHs massive enough to emit GWs in the PTA band, gas is thought to be scarce, and is therefore unlikely to drive SMBH binaries close enough to become GW dominated.  Furthermore, once the SMBH binaries reach orbital separations of $\mathcal{O}$(1 pc), simple estimates imply that there are not enough stars available in the region of phase space capable of removing the necessary amount of energy and angular momentum for the SMBH binary to merge via GW emission within a Hubble time.  This region of phase space for stars is referred to as the loss cone, and the depletion of the loss cone is often referred to as the last parsec problem~\cite{2003AIPC..686..201M}.  There are many proposed resolutions to the last parsec problem, but they primarily involve mechanisms that can refill the loss cone~(see \cite{Merritt:2005LR} and references therein).  

It has commonly been assumed in PTA analyses that the last parsec problem is solved by some unspecified mechanism, but that GWs are the dominant driver of binary mergers throughout the frequency range ($\sim10^{-9} - 10^{-6}$ Hz) to which pulsar timing is sensitive. This is not, however, guaranteed to be the case, and it is important to remember that the SMBH binaries of interest emit GWs throughout their evolution, even when the driving evolutionary force is something other than GWs. This means that the stochastic background of GWs to which PTAs will be sensitive could have a spectral shape that is significantly different than that predicted by assuming purely GW-driven binary dynamics~\cite{2004ApJ...611..623S, 2011MNRAS.411.1467K,2012arXiv1211.4590M,0004-637X-789-2-156,2013MNRAS.433L...1S,Shannon:2013wma,2014MNRAS.442...56R}. 

This reality can have both negative and positive consequences. The negative effect of gas- or stellar-driven binary evolution is that it causes the amplitude of the stochastic GW background within the PTA sensitivity band to be lower than what it would be if all mergers were GW-driven throughout the band. However, the presence of more complicated spectra, which could be shaped by many astrophysical processes, would give us the opportunity to learn about the galactic environments in which SMBH binaries evolve and merge. 

In this paper we explore both of these possibilities, using the techniques of Bayesian inference and model selection. First, in Section~\ref{Sec:Specmodels} we discuss some of the processes that can affect the shape of the stochastic spectrum, and how to determine what that shape will be. In Section~\ref{Sec:results}, we describe our simulations and the analysis techniques used, and present the results of our analysis. We find that the detection of a stochastic GW background is not hindered, and can in fact be significantly helped, by using a more complicated model for the spectrum that includes the possibility of non-GW inspiral mechanisms. We also show the extent to which the parameters that describe this spectrum can be measured in the event of a detection, and constrained in the event of a non-detection. Throughout this section we analyze the effects that the choice of priors has on model selection and parameter estimation, and discuss which prior choice is appropriate for a particular goal. Finally, we summarize and conclude in Section~\ref{Sec:summary}.  Unless otherwise noted, we will work in geometrized units where Newton's constant $G$ and the speed of light $c$ satisfy $G = c = 1$.

\section{Spectral Models}
\label{Sec:Specmodels}
The characteristic amplitude of the stochastic GW signal generated by a population of inspiralling SMBH binaries on circular orbits can be calculated via~\cite{Phinney:2001di,2008MNRAS.390..192S,0004-637X-789-2-156}
\begin{equation}
h_c^2(f) = \int_0^{\infty}dz\int_0^{\infty}d\mathcal{M} \frac{d^3N}{dz\,d\mathcal{M}\,dt}\frac{dt}{d\ln f} h^2(f)\,.
\label{Eq:fullspectrum}
\end{equation}
In this expression, $h(f)$ is the instantaneous GW strain amplitude emitted by a single circular binary with a Keplerian rest frame orbital frequency of $f/2$, since circular binaries emit gravitational waves at twice
the orbital frequency, and is given by \cite{Thorne300}
\begin{equation}
h(f)=2(4\pi)^{1/3}\frac{f^{2/3}\mathcal{M} ^{5/3}}{D_L}\,
\end{equation}
where $D_L$ is the luminosity distance to the source. For a binary with component masses $M_1$ and $M_2$, $\mathcal{M} \equiv (M_1M_2)^{3/5}(M_1+M_2)^{2/5}$ is a combination of masses known as the chirp mass, $z$ is the redshift, and $t$ is time measured in the binary rest frame.
We note that the amplitude $h(f)$ applies regardless of whether the source is
evolving primarily under the influence of gravitational waves, or is dominated by some other dynamical effect. The term 
\begin{equation}
\frac{d^3N}{dz\,d\mathcal{M}\,dt}\frac{dt}{d\ln f} = \frac{d^3N}{dz\,d\mathcal{M}\,d\ln f}
\end{equation}
is the differential number of inspiraling binaries per unit $\mathcal{M}$, $z$, and $\ln f$. The term $dt/d\ln f$ encodes the frequency evolution of the binary. For a population of circular binaries that are driven purely by GW emission, this frequency evolution is given by the standard quadrupole formula:
\begin{equation}
\frac{d t}{d \ln f} = \frac{5}{64 \pi^{8/3}} \mathcal{M} ^{-5/3} f^{-8/3}\,.
\label{Eq:dtdlnf}
\end{equation}
With this substitution, Eq.~\ref{Eq:fullspectrum} becomes~\cite{Phinney:2001di}
\begin{equation}
h_c^2 (f) = \frac{4 f^{-4/3}}{3 \pi^{1/3}} \int \int dz d\mathcal{M} \frac{d^2 n }{dz d\mathcal{M}} \frac{\mathcal{M}^{5/3}}{(1+z)^{1/3}},
\label{Eq:stdspectrum}
\end{equation}
where $dn^2/(dz d\mathcal{M})\equiv dN^3/(dV_c\, d\mathcal{M} \, dz)$ is the comoving number density of merged remnants per unit $\mathcal{M}$ and $z$, with $V_c$ being the comoving volume. Finally, $dV_c$ intervals are related to $dz$ intervals via~\cite{Phinney:2001di}
\be
dV_c = \frac{4\,\pi \,D_L^2}{H_{\rm o}(1+z)^2\sqrt{\Omega_M(1+z)^3+\Omega_{\Lambda}}}dz\,,
\ee
where $H_{\rm o}$, $\Omega_M$, and $\Omega_{\Lambda}$ are the Hubble constant, total (baryon and dark matter) mass-energy density, and dark energy density, respectively, and $D_L$ is related to $z$ in a flat cosmology via
\be
D_L(z)=\frac{1+z}{H_{\rm o}} \dint_{0}^{z} \frac{dz'}{\sqrt{\Omega_M(1+z')^3+\Omega_{\Lambda}}}
\ee
(see \cite{Hogg} and references therein).

Eq.~(\ref{Eq:stdspectrum}) is often re-parameterized as $h_c = A(f/\rm{yr}^{-1})^{-2/3}$, where $A$ is the amplitude at $f = 1 \rm{yr}^{-1}$. $A$ is therefore the standard metric that has been used throughout the literature for characterizing the sensitivity of PTAs, although this only makes sense if we assume all binaries have circular orbits and are gravitational-wave dominated. 

GWs cannot, however, be the only driver of SMBH coalescence - not if these binaries are to merge within a Hubble time. There are many other mechanisms, such as interactions between the SMBH binary and surrounding stars and gas, that draw the binaries together after their galaxies merge. These mechanisms will alter the calculation above exclusively via a change in $dt/d\ln f$; generically, $dt/d\ln f$ can be no smaller than the prediction of Eq.~\ref{Eq:dtdlnf}, since any other dominating mechanism will cause binaries to evolve more rapidly through frequency. This will result in a different frequency dependence in the final power spectrum, such that the spectral slope will be smaller at frequencies dominated by other mechanisms. 

To see how other dynamical effects will impact the shape of the spectrum, we can recast Eq.~(\ref{Eq:fullspectrum}) as
\be
h_c^2(f) \sim f \frac{dN}{df} h^2(f) \sim \frac{f^{7/3}}{\frac{df}{dt}}.
\label{Eq:pspec}
\ee
Then we can write
\be
\frac{df}{dt} = \sum_i \left( \frac{df}{dt}\right)_i,
\label{Eq:sum}
\ee
where the sum extends over all the physical processes that drive the black holes to inspiral. For instance, for GW-driven evolution
\be
\left(\frac{df}{dt} \right)_{\rm{GW}} \sim f^{11/3}.
\ee
Given the strong frequency dependence of GW-driven inspiral, we will assume that at least the final stage of binary evolution is dominated by GW emission, and pull this term out of the sum in Eq.~\ref{Eq:sum}. We can then recast the expression in Eq.~(\ref{Eq:pspec}) in the form
\be
h_c(f) = A \frac{(f/f_{\rm year})^{-2/3}}{(1+ \sum_i a_i \, (f/f_{\rm year})^{(\gamma_i - 11/3)})^{1/2}},
\ee
where $f_{\rm year}$ is the fiducial frequency of $1/\rm{year}$, the $\gamma_i$'s encode the frequency dependence introduced by a particular astrophysical effect, the $a_i$ are the dimensionless relative amplitudes of the astrophysical processes driving the binary toward merger scaled relative to the GW driven case at $f=f_{\rm year}$, and $A$ is a dimensionless amplitude. Note that unless the $a_i=0$, $A$ is {\em not} equal to the characteristic amplitude at a frequency of $1/{\rm year}$. On the other hand, for most scenarios we expect the $a_i \ll 1$, so to a good approximation, $A$ is close to equaling $h_c(f=f_{\rm year})$.

Before we consider specific alternative dynamical effects to GW-driven inspiral, it is useful to briefly review our understanding of the evolutionary histories of SMBH binaries. The early SMBH dynamics within merging galaxies is driven by dynamical friction \cite{Chandrasekhar:1943ys} wherein each SMBH and its concomitant stellar bulge and dark matter subhalo gravitationally focus first the dark matter and then the baryonic content of the companion galaxy.  This focusing of excess mass behind the SMBH induces a drag, thereby causing the SMBH to drop further into the gravitational potential well of the merging pair.  Dynamical friction continues to operate until the SMBH orbital velocity exceeds the stellar velocity dispersion, at which point the binary is said to have hardened.  Subsequent evolution of the binary is driven by stellar scattering, in which background stars on centrophilic orbits (i.~e.~stars outside the SMBH binary that are scattered onto low angular-momentum orbits toward the center of the gravitational potential) interact with the SMBH binary in a three-body interaction, and remove energy and angular momentum from the SMBH binary as a result~\cite{Quinlan:1996vp}. The last parsec problem occurs when there are insufficient stars on the necessary orbits to continue driving the SMBH binary toward the GW-dominated regime.  There are several proposed solutions to this problem, including any form of asymmetry in the gravitational potential (see \cite{2013ApJ...773..100K} and references therein), but all of these solutions can be considered mechanisms to refill the loss cone, so that stellar-induced hardening via three-body interaction can continue until GWs are able to efficiently drive the binary onward to merging. 

The semi-major axis $a$ of a SMBH binary that interacts with a background of stars via three-body scattering evolves according to~\cite{Quinlan:1996vp}
\begin{equation}
\frac{da}{dt} = \frac{a^2 \rho}{\sigma} H\,,
\label{Eq:dadtstars}
\end{equation}
where $\rho$ is the density of background stars and $\sigma$ is their velocity dispersion, and $H$ is approximately constant and was estimated in \cite{Quinlan:1996vp} via N-body simulations. Thus, for stellar scattering $df/dt \sim f^{1/3}$.
If we assume that a binary is driven by stellar scattering immediately prior to being driven by the emission of GWs, we can find the frequency at which GWs begin to dominate by equating $(da/dt)_{\rm{stars}}$ to from Eq.~\ref{Eq:dadtstars} to $(da/dt)_{\rm{GW}}$ under the influence of gravitational-wave emission.
This crossover frequency has been estimated as $f_b \approx 10^{-7} M_6^{-7/10} q^{-3/10}$ Hz, where $M_6$ is the total mass of the binary in millions of solar masses~\cite{Quinlan:1996vp}.
In what follows, we will call such transition frequencies ``bend frequencies'', since the slope of $h_c(f)$ changes as one hardening process takes over from another. For a two part $da/dt$ model (or, equivalently, $df/dt$), where some other dynamical mechanism
gives way to GW-driven evolution, we have
\begin{equation}\label{hc}
h_c(f) = A \frac{(f/f_{\rm year})^{-2/3}}{(1 + (f_b/f)^\kappa)^{1/2}}\,,
\end{equation}
with $\kappa = 10/3$ being the appropriate value for the specific case of stellar scattering. We must mention that stellar scattering is not the only mechanism that could potentially influence the inspiral of SMBH binaries. Other possible effects include torquing of the binary from a circumbinary disks, where the details of the effect depend sensitively on the dissipative physics of the disk, and gas-driven evolution due to massive inflows of gas that can be triggered by dynamical instabilities during the galactic merger~\cite{Sesana:2014wta,2013MNRAS.433L...1S,Sesana:2013dma,2012AdAst2012E...3D,2014MNRAS.442...56R,Shannon:2013wma,Jaffe:2002rt,2010ApJ...719..851S,Enoki:2006kj}. In general, a particular physical effect will lead to an expression for $da/dt$, which can be mapped to values of $f_b$ and $\kappa$ in the GW spectrum. Given the presumed absence of gas in the massive elliptical galaxies that host PTA sources, disk- or gas-driven inspiral is thought to be less probable than stellar scattering, so we will focus primarily on the latter as the dominant effect preceding GW-driven inspiral.  We also mention that, even in the case of stellar-driven inspiral, it is possible that the balance of energy and angular momentum removal could drive SMBH binaries to nonzero eccentricities when they transition to being GW dominated \cite{Quinlan:1996vp}.  Having said this, zero eccentricity appears to be an attractor solution for binaries in numerical experiments with plausible initial conditions of the stellar population \cite{Quinlan:1996vp}, so we will focus on circular binaries hereafter.

Realistic astrophysical binaries will be driven by multiple mechanisms throughout their evolutionary history, which is reflected in the summation in Eq.~(\ref{Eq:sum}). This will result in a GW power spectrum with several slopes and several transition frequencies. It is likely, however, that in order to measure the parameters of such a complicated power spectrum, it will be necessary to make a high signal to noise ratio (SNR) detection of the stochastic GW background. In the early days of PTA data analysis, a simple model that allows for a single transition between GW-dominated evolution and some other power law will most likely suffice. 

\begin{figure}
\includegraphics[clip=true,width=0.48\textwidth]{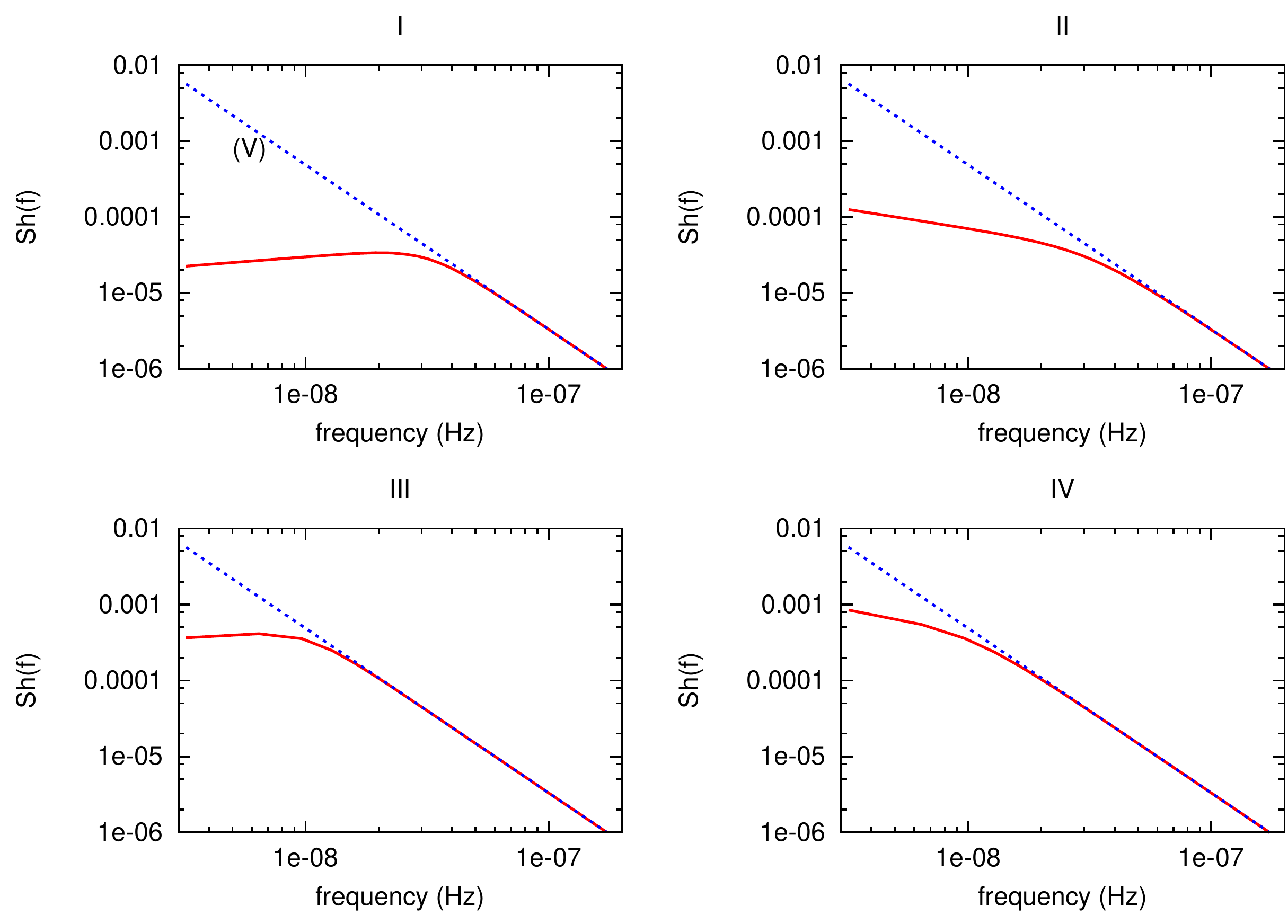} 
\caption{\label{fig:fourspec} Examples of the five different spectral shapes we will investigate. (I) has $f_b = 10^{-8}$ Hz and $\kappa = 29/6$, (II) has $f_b = 10^{-8}$ Hz and $\kappa = 10/3$, (III) has $f_b = 3\times10^{-8}$ Hz and $\kappa = 29/6$, (IV) has $f_b = 3\times10^{-8}$ Hz and $\kappa = 10/3$, and (V) is the signal from binaries that are driven purely by GWs throughout the band.}
\end{figure}

Figure~\ref{fig:fourspec} shows examples of the different types of simulated signals we will analyze in Section~\ref{Sec:results}. Plotted in this figure are the four basic shapes of spectra which we will investigate, along with the baseline spectrum of a purely GW-driven binary population. The spectra include two different bend frequencies, $f_b = 1\times 10^{-8}$ Hz and $f_b = 3\times 10^{-8}$ Hz, and two different slopes for the low-frequency part of the data, $\kappa = 10/3$, which corresponds to stellar scattering, and $\kappa = 26/9$, which is chosen to be a somewhat more dramatic bend. These different shapes are labeled with roman numerals. Further details of the simulated data will be discussed in the next section. 

The final astrophysical function that enters into Eq.~\eqref{hc}, is the merger rate, and it is this rate that determines the overall amplitude of the GW background. When performing a Bayesian analysis with the spectral model \eqref{hc}, it is necessary to provide priors on all the parameters. A prior on a particular parameter encodes any information that is known, or believed, \emph{a priori} about that parameter.
In the analysis presented in the next section, we will focus on two different astrophysically motivated priors on $A$, as well as priors that are uniform in $A$ and in $\ln A$. The two astrophysical priors are motivated by the work of McWilliams {\it et al.}~\cite{2012arXiv1211.4590M,0004-637X-789-2-156} (Model A), and Sesana and Ravi {\it et al.}~\cite{2013MNRAS.433L...1S,2014MNRAS.442...56R} (Model B). Both models can be approximated as Gaussian distributions in $\ln A$.
The prior for Model A is centered at $A=4.1\times10^{-15}$, with an uncertainty in $\ln A$ of $\sigma = 0.6$ \cite{0004-637X-789-2-156}. The Model B prior is centered at $A=10^{-15}$, with an uncertainty in $\ln A$ of $\sigma = 0.5$. Since these prior distributions are Gaussian, neither requires the specification of an upper or lower limit on $A$. The third prior that we consider, which is uniform in $A$, has the natural lower limit of $A=0$. To calculate the upper limit for this prior, we impose the (rather conservative) constraint that the energy density in GWs in the PTA observing band cannot close the universe. An alternative would be to demand that the energy density in GWs from PTA sources cannot close the universe, which would extend the integration over the energy density to include the out-of-band merger and ringdown of the black hole binaries. We chose to use the frequency
range covered by the PTA observing band since those are the frequencies that are directly constrained by the data.  To use this requirement to impose an upper limit on $A$ we relate $h_c$ to the energy density per
logarithmic frequency interval via
\begin{equation}
\Omega_{\rm{GW}} = \frac{4\pi^2}{3 H_0^2} f^2 h_c^2(f)\,,
\end{equation}
then impose that
\begin{equation}
\int_{f_{\rm{min}}}^{f_{\rm{max}}} \Omega_{\rm{GW}} (f) \;  d\ln f  < 1.
\end{equation}
The lower frequency limit of this integral is unimportant since the integral is dominated by the high frequency behavior. Performing the integral results in a limit on $A$ of the form
\begin{equation}
A^2 < f_y ^{-4/3} f_{\rm{max}}^{-2/3} \frac{H_0^2}{2 \pi^2}.
\end{equation}
For $f_{\rm{max}} = 2.0\times10^{-6}$ Hz, this implies that $A < 4.4\times10^{-12}$, which is the upper limit we use when employing a uniform prior in $A$ throughout our analyses. We take the uniform prior on
the logarithm of the amplitude to share the same upper limit derived for the uniform amplitude prior, but there is no natural value for the lower limit. This prevents us from using this prior to set upper
limits, but by choosing a lower limit several orders of magnitude below the noise level, it is possible to use the uniform logarithmic prior for parameter estimation once a detection has been made.

The other parameters that enter into our model for the GW spectrum are $f_b$ and $\kappa$, as discussed above. The prior we choose to employ on $f_b$ is uniform in $\ln(f_b)$, indicating our lack of knowledge about even the order of magnitude of the frequency at which the GW slope may bend, and has a lower limit of $f_b = 2.5\times10^{-9}$ Hz (which is below the lower edge of the sensitivity band for a 10 year data span), and an upper limit of $f_b = 10^{-7}$ Hz. This upper limit is somewhat arbitrary, but it does indicate our belief that GWs must, at some high enough frequency, dominate the evolution of SMBH binaries. For $\kappa$, we use a prior that is uniform in $\kappa$, with a range of $0$ to $23/3$ (the non-integer value for the upper end of the range is due to us changing our conventions for the parameterization of the spectrum late in the project).

We note that we have chosen to study our ability to measure the low-frequency slope, bend frequency, and amplitude for a set of simulated data, rather than trying to assess how efficiently or frequently the last parsec problem is solved, or what the correct correlation is between the black hole mass and a particular property of the host galaxy.  Our principle motivation for this choice is that the astrophysically-motivated distributions only differ by a factor of $\sim2$--$3$ in predicted signal amplitude between the most conservative and the most optimistic estimates, despite making different assumptions about the solution to the last parsec problem and the correlation between black-hole mass and host-galaxy properties.  Furthermore, there are multiple elements that contribute to this level of amplitude uncertainty, including the overall galaxy merger rate and its dependence on mass and environment, in addition to the last parsec solution and the choice of black-hole mass - host-galaxy-property correlation.  Given our comparative ignorance of the low-frequency signal slope and the transition frequency between the dynamical process dominating the final parsec and GW-driven evolution, we choose to focus our study on constraining these parameters for a given amplitude.

Another approach we could have taken with the amplitude prior is to separate the uncertainties into an observational part (from factors such as the rate of galaxy mergers and the $M$-$\sigma$ relation) and a
theoretical part (from factors such as the efficiency of dynamical friction in hardening the binary and the fraction of systems where the last parsec problem is overcome). This could be done by writing the prior as a
Gaussian distribution in $\ln A$, centered at some value $\ln \bar{A}$, with width $\sigma_{\rm obs}$ to account for the observational uncertainties. The central value, $\ln \bar{A}$, would then be a hyper-parameter
to be determined by the data. If we quantify the uncertainty on the central value $\bar{A}$ as $\bar{A} = \eta A_*$, where $A_*$ is some reference value and $\eta$ encodes the theoretical uncertainty in the merger efficiency,
then assigning a Gaussian prior on the hyper-parameter $\ln \eta$ of width $\sigma_{\rm th}$ centered on $\eta_*$, and marginalizing over the hyper parameter,  leads to Gaussian distributions of the form used for
Models A and B with width $\sigma = \sqrt{\sigma_{\rm obs}^2+\sigma_{\rm th}^2}$. Alternatively, a uniform prior on $\ln \eta$ over some range leads to a roughly uniform prior on $\ln A$ between
$\ln (\eta_{\rm min} A_*)$ and  $\ln (\eta_{\rm max} A_*)$ with rounded edges of width $\sigma_{\rm ons}$.  Thus, the analyses we consider are equivalent to a hierarchical Bayesian analysis that
separates out the theoretical and observational uncertainties for certain choices of the hyper-prior.

\section{Simulated Data and Analysis}
\label{Sec:results}
Our simulated data set consists of the timing residuals from 20 pulsars, randomly distributed on the sky, and observed for 10 years with a two-week cadence. The data from each pulsar is generated including white noise at a level of $200$ ns, and no red noise. We then recover this simulated signal using a model that includes the three parameters that describe the GW spectrum, ${A, f_b, \kappa}$, and the noise parameters for each pulsar. These include a white noise level, red noise level, and red noise slope. The white noise is fully described by the amplitude of its power spectral density (PSD), which we label $S_n$. The prior on $S_n$ was uniform in $\ln (S_n)$, and ranged from $S_n = 4 \times10^{-18} \, \rm{Hz}^{-1/2}$ to $S_n =  10^{-2} \, \rm{Hz}^{-1/2}$. The red noise is parameterized by its PSD amplitude, $S_r$, and by a slope, $r$, as $S_r (f) = S_r (f/f_{\rm{year}})^r$. The prior on the red noise amplitude was also uniform in $\ln (S_r)$, with the same range as $S_n$, and the prior on the slope $r$ was uniform from $-2$ to $-6$.

Finally, we include two parameters for each pulsar that encode the effects of the timing model on the timing residuals. As discussed briefly in the introduction, the timing model used to predict the TOA of radio pulses from a given pulsar includes a large set of parameters~\cite{lorimer2005handbook,Hobbs21122012,2010ApJ...720L.201C}. In this analysis we only consider the two timing model parameters that have the greatest effect on the low-frequency sensitivity. These are the quadratic and linear terms in the spin down model for each pulsar, which take the form
\begin{equation}
P(t) = P_0 + \dot{P}_0 t + \ddot{P}_0 t^2 + \ldots.
\end{equation} 
Here, $P_0$ is the initial spin period of the pulsar, and $\dot{P}_0$ and $\ddot{P}_0$ encode the way that this spin evolves in time.

In order to speed up our analysis, we choose to perform our calculations in the Fourier domain. We therefore need to understand how the quadratic and linear terms in the timing model translate to effects on the timing residuals as a function of frequency. The Fourier transform of the timing model is given by
\begin{equation}
\tilde{P} (f_k) = \int_0^T P(t) e^{i 2 \pi f_k t} dt,
\end{equation}
where $f_k = k/T$ for integer $k$, and $T$ is the observation time. This integral evaluates to
\begin{equation}
\tilde{P}(f_k) = T P_0 \delta_{k0} - \frac{i \dot{P}_0 T^2}{2 \pi k} - \frac{i \ddot{P}_0 T^3}{2 \pi k} + \frac{\ddot{P}_0 T^3}{2 \pi^2 k^2} + \ldots
\end{equation}
The $k=0$ term is simply a constant offset that we can ignore. Writing $a = \ddot{P}_0 T^3/(2\pi^2)$ and $b = -(\dot{P}_0 T^2 + \ddot{P}_0 T^3)/(2\pi)$, the timing model for each pulsar can be written as
\begin{equation}
\tilde{P}_k = \frac{a}{k^2} + \frac{i b}{k}.
\end{equation}
A model of this form, with independent $a$ and $b$ for each pulsar, is then subtracted from the TOAs.

The full set of parameters in our model thus consists of the five noise/timing parameters for each pulsar, and three parameters to describe the GW background - ${A, f_b, \kappa}$. With this set of parameters, the likelihood is defined by~\cite{Cornish:2012cr}
\begin{equation}
p(d|s,n) = \frac{1}{\sqrt{(2\pi)^L \det C}} \exp \left(-\frac{1}{2} \sum_{ab} \sum_{ij} r_{ai} C^{-1}_{(ai)(bj)} r_{bj}  \right),
\end{equation}
where C is the covariance matrix, which depends on both the noise in the individual pulsars and on the GW background, and $r= d-s$ denotes the timing residuals after the subtraction of the timing model $s$ from the data $d$. The indices $a$ and $b$ label individual pulsars, and run from $1$ to the number of pulsars, $N_p$. The indices $i$ and $j$ label the data samples, i.e. individual frequency bins. Since our simulated data is stationary, the
correlation matrix is diagonal in $i,j$ and $C_{(ai)(bj)} \rightarrow C_{ab}(f)$.

The timing model parameters for each pulsar enter the likelihood in the timing residuals; they are subtracted from the TOAs before the likelihood is evaluated. The red and white noise contributions for each pulsar enter along the diagonal of the covariance matrix. Finally, the GW signal enters via the Hellings and Downs~\cite{HellingsDowns} correlation matrix, which has the form
\begin{equation}
H_{ab} = \frac{1}{2} + \frac{3(1 - \cos \theta_{ab})}{4}\ln \left(\frac{1-\cos\theta_{ab}}{2} \right) - \frac{1- \cos \theta_{ab}}{8} + \frac{1}{2} \delta_{ab}.
\end{equation}
The full covariance matrix is then given by
\begin{equation}
C_{ab} (f) = S_h(f) \frac{H_{ab} }{3}+ \delta_{ab} \left\{S_{n_a} + S_{r_a} (f/f_y)^{r_a}\right\},
\end{equation}
where $S_h(f)$ is the PSD of the GW background, $S_{n_a}$ is the PSD of the white noise, $S_{r_a}$ is the amplitude of the PSD of the red noise, and $r_a$ is the slope of the red noise.

\subsection{Methods}
Our analysis is carried out within the framework of Bayesian inference, using the technique of Markov chain Monte Carlo (MCMC). To calculate Bayes factors\footnote{Given equal prior belief in the validity of two models, A and B, the Bayes factor, $B_{AB}$, between models A and B, given the observed data, is the `betting odds' that model A is the better theory, rather than model B.}, we must calculate the evidence for each model, which necessitates performing an integral over the full parameter space. For this integral, we use the technique of thermodynamic integration~\cite{PhysRevD.80.063007, Cornish:2014kda, Littenberg:2014oda}. This technique necessitates the use of parallel tempering, in which multiple chains are run at different `temperatures,' which are defined by changing the likelihood to
\begin{equation}
p(d|s,n,\beta) = p(d|s,n)^{\beta},
\end{equation}
where $\beta$ is analogous to an inverse temperature. This effectively `softens' the likelihood, allowing the chains to effectively sample the full posterior. Chains with different temperatures are allowed to swap parameters
with a probability given by the Hastings ratio
\begin{equation}
H_{i\leftrightarrow j} = {\rm min} \left(\frac{p(d|s_i,n_i,\beta_j) p(d|s_j,n_j,\beta_i)}{p(d|s_i,n_i,\beta_i) p(d|s_j,n_j,\beta_j)}, 1\right).
\end{equation}
The evidence for each model is then given by
\be
\ln p(d) = \int_0^i E_\beta [\ln p(d|\vec{x})] d\beta,
\ee
where $E_\beta$ denotes the expectation at inverse temperature $\beta$. Given equal prior belief in two models, the Bayes factor is then simply the evidence ratio between them.
The technique of parallel tempering is useful not only as a means of calculating Bayes factors, but as a tool for improving the convergence of the MCMC runs. 

In order to further speed convergence, we use a cocktail of jump proposals. These include small random-walk jumps of varying sizes, which we define as draws in each parameter from a Gaussian centered at the current location; parallel tempering swaps; and differential evolution proposals~\cite{DiffEv}. We find that the differential evolution jumps are particularly helpful in encouraging efficient exploration.

\subsection{Detection and Model Selection}
\begin{table}
\centering
\begin{tabular}{ c  c  c  c  c  }
\hline
      Type    & sub-Type & $A\times10^{15}$ & $f_b$ (Hz) & $\kappa$\\
  \hline  
  \hline                     
      I      	& a & 0.08 & $3\times10^{-8}$ & 29/6\\
            	& b & 2.0    &  $3\times10^{-8}$              & 29/6\\
            	& c & 4.0    &   $3\times10^{-8}$              & 29/6\\
       II     	& a & 0.08 & $3\times10^{-8}$ & 10/3\\
           	& b & 2.0    &   $3\times10^{-8}$              & 10/3\\
           	& c & 4.0    &    $3\times10^{-8}$             & 10/3\\
       III      & a & 0.08 & $10^{-8}$ & 29/6\\
           	& b & 2.0    &  $10^{-8}$              & 29/6\\
           	& c & 4.0    &  $10^{-8}$               & 29/6\\
       IV      & a & 0.08 & $10^{-8}$ & 10/3\\
           	& b & 2.0    &  $10^{-8}$               & 10/3\\
           	& c & 4.0    &  $10^{-8}$               & 10/3\\
       V    	 & a & 0.08 & $0$ 	& N/A\\
           	& b & 2.0    &    $0$  & N/A \\
           	& c & 4.0    &      $0$          & N/A\\
  \hline  
\end{tabular}
 \caption{The different types of simulated signals for Figures~\ref{fig:BFflat} and ~\ref{fig:BFall}.}
 \label{table:models}
 \end{table}

\begin{figure}
\includegraphics[clip=true,angle=0,width=0.48\textwidth]{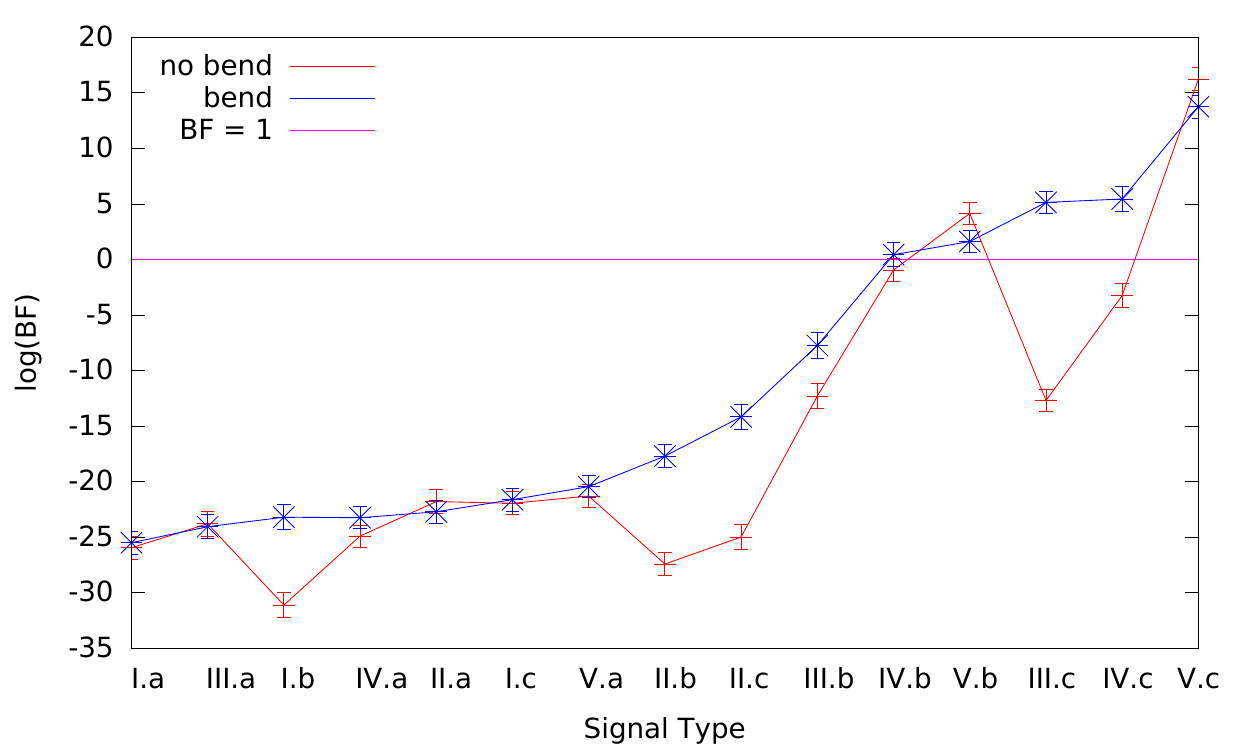} 
\caption{\label{fig:BFflat} The Bayes factors between noise and two GW signal models, both with uniform amplitude prior, one with a bend in the spectrum, the other assuming a purely GW driven evolution.
A Bayes factor larger than unity indicates a detectable GW background. Unsurprisingly, the model that allows for a bend in the spectrum fares much better than the GW-driven model in cases where the
simulated signal spectrum includes a bend (types I,II,III and IV), and performs only slightly worse when the simulated signal is purely GW-driven (type V).}
\end{figure}
\begin{figure}
\includegraphics[clip=true,angle=0,width=0.48\textwidth]{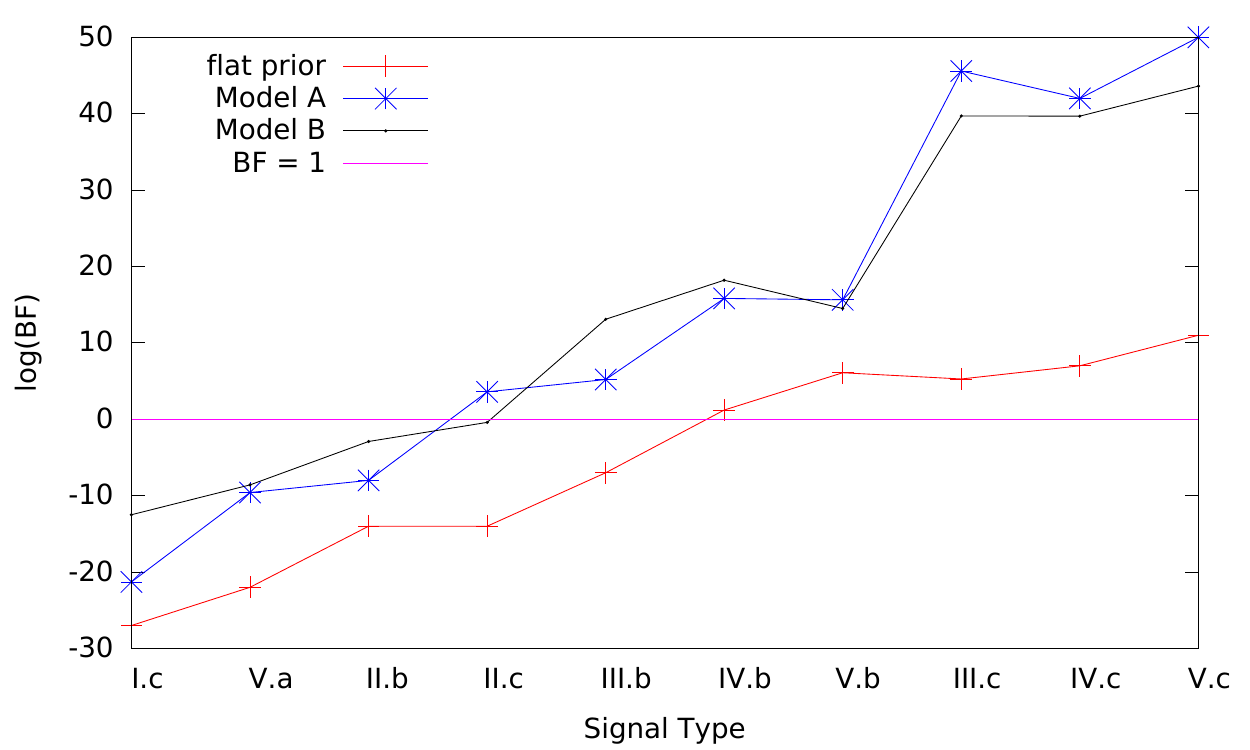} 
\caption{\label{fig:BFall} The Bayes factors between GW signal models with three different priors - uniform in amplitude and the Model A and B astrophysical amplitude priors - and noise. In this
study, all the GW models allowed for a bend in the spectrum. A Bayes factor larger than unity indicates a detectable GW background. Here we see that both astrophysical priors outperform the
uniform prior for detectability, and also that we are able to distinguish between the two astrophysical models.}
\end{figure}

We first investigate the impact on detectability of the shape of the GW spectrum, both in the simulated signal and in the model used to extract it. We do this by simulating GW backgrounds of varying shapes and amplitudes, described in Table~\ref{table:models}, and recovering these signals using templates that either do or do not include a bend in the spectrum. The templates that do not include a bend are consistent with the belief that the SMBH binary population is driven purely by GW emission throughout the sensitivity band. Figure~\ref{fig:BFflat} shows the results of this study. The prior on the GW amplitude for the two signal models (with a bend and without)  was uniform in amplitude. The horizontal axis is labeled by the simulated signals, which have been arranged from left to right in ascending order of detectability for the model that includes a bend. The vertical axis shows the Bayes factor in favor of either signal model, where a Bayes factor larger than unity indicates detectability.  There are two main features to notice in Figure~\ref{fig:BFflat}. The first is that the model with no bend, i.e. for which the SMBH binaries are driven purely by GW emission throughout, does not perform appreciably better than the more complicated model that includes a bend for any simulated signal. This is true even when the signals themselves do not include a bend in the spectrum. This tells us that using the more complicated model for the purpose of detection will not significantly decrease our sensitivity to a GW background\footnote{The best approach would be to marginalize over the model dimension along with the model parameters using a Reversible Jump Markov Chain Monte Carlo algorithm so that the optimal model is selected by the data and not guessed in advance.}. The second, related feature is that the Bayes factor for the model that allows for a bend is consistently favored over the no-bend model when the simulated signal has a bend. From this we can see that we will be able to perform model selection between astrophysical models that predict a slope change in the spectrum, and those that predict a purely GW-driven evolution across the PTA frequency band.

The other main prediction of astrophysical models is reflected in the prior we impose on the amplitude of the GW background. In the previous section, we discussed the different amplitude priors from the merger models of McWilliams {\it et al.}~\cite{0004-637X-789-2-156} (Model A) and Sesana and Ravi {\it et al.}~\cite{2013MNRAS.433L...1S,2014MNRAS.442...56R} (Model B). In Figure~\ref{fig:BFall}, we demonstrate the ability to perform model selection between these two priors. This figure again shows the Bayes factors between a set of GW spectrum models and noise. We see that both of the astrophysical models outperform the uniform amplitude prior, and that the Model A and Model B priors are distinguishable from each other, with Model A preferred for large amplitude signals, and Model B preferred for small amplitude signals.

\subsection{Parameter Estimation}
\begin{figure*}[htp]
\includegraphics[clip=true,angle=0,width=0.48\textwidth]{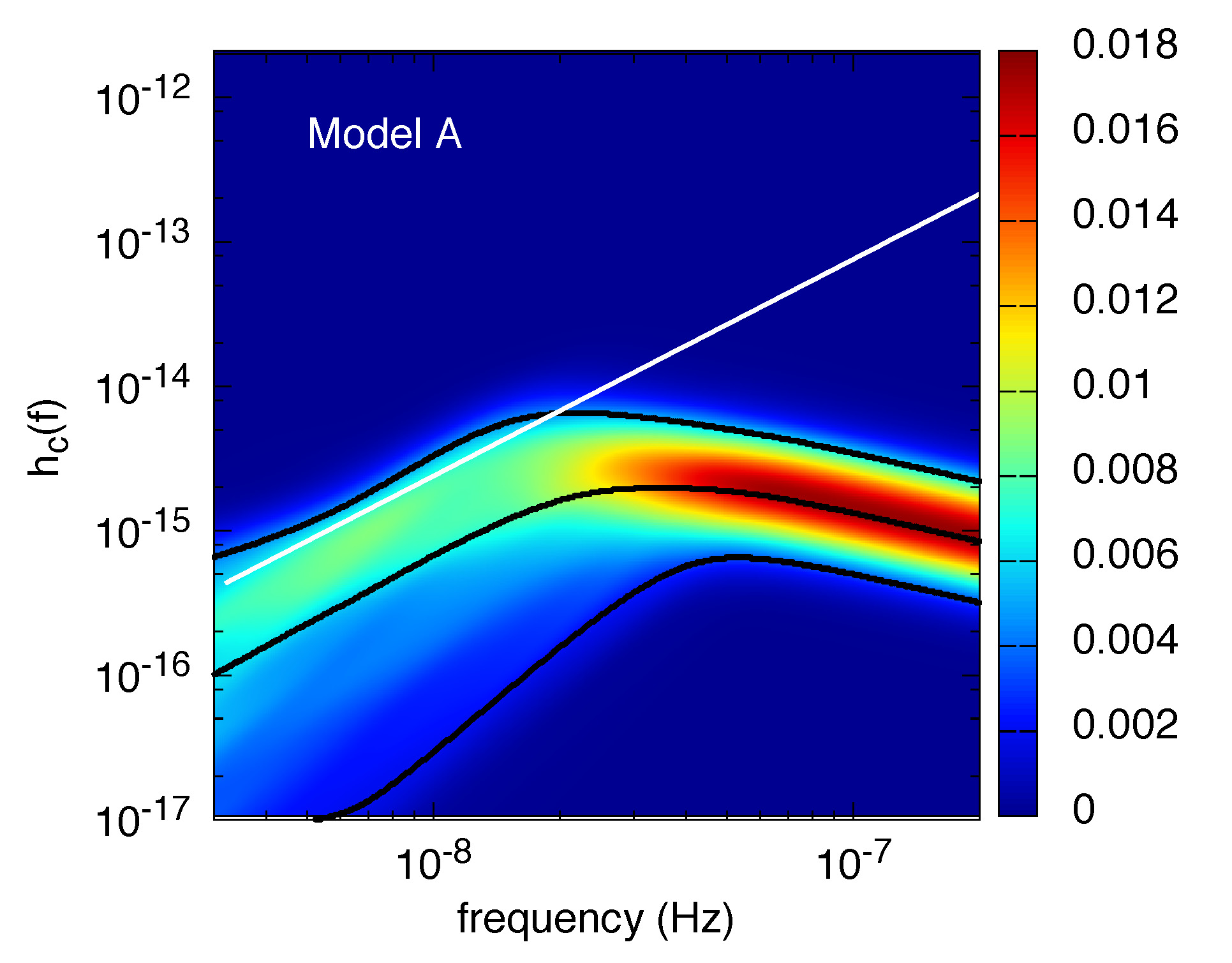} 
\includegraphics[clip=true,angle=0,width=0.48\textwidth]{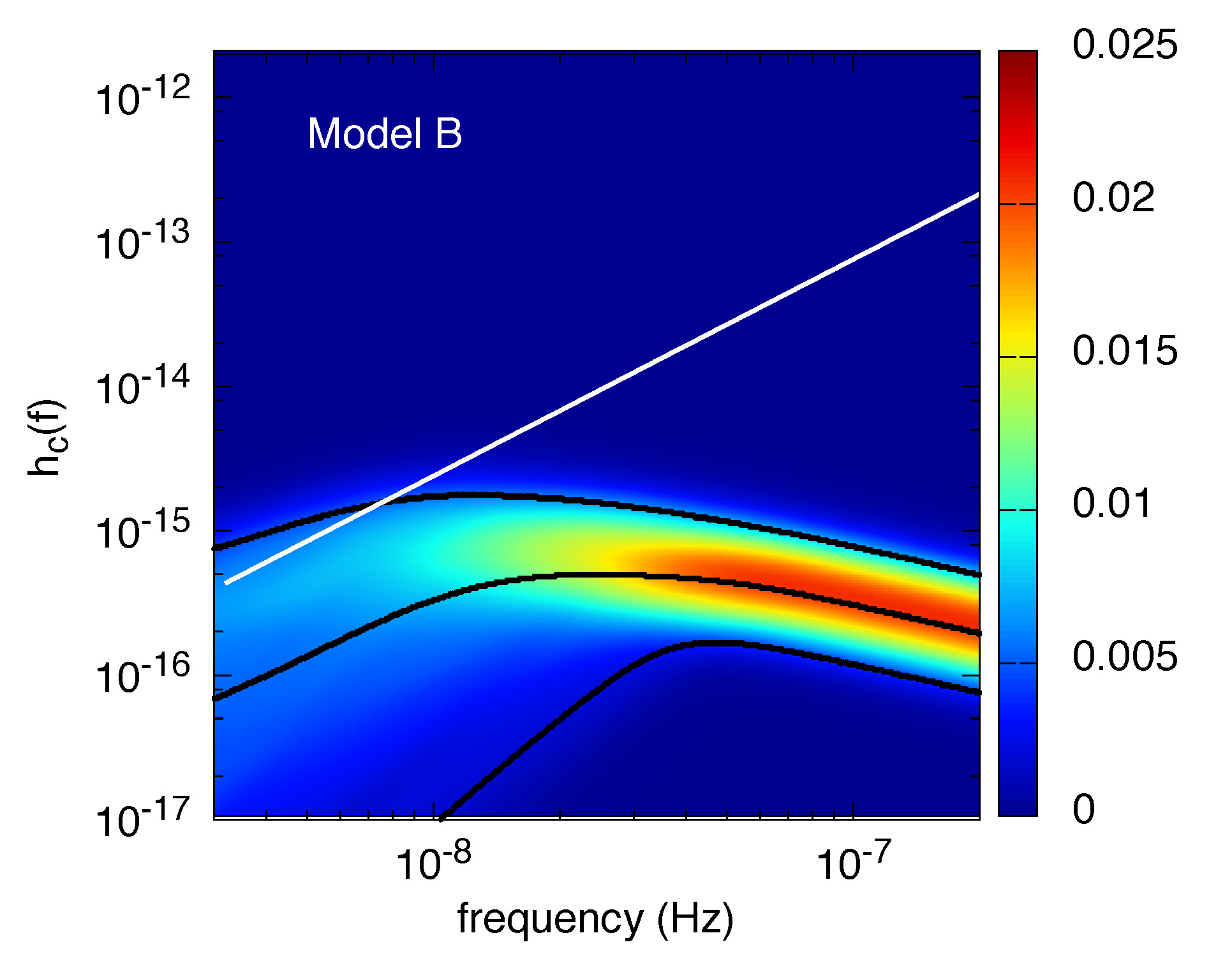} 
\includegraphics[clip=true,angle=0,width=0.48\textwidth]{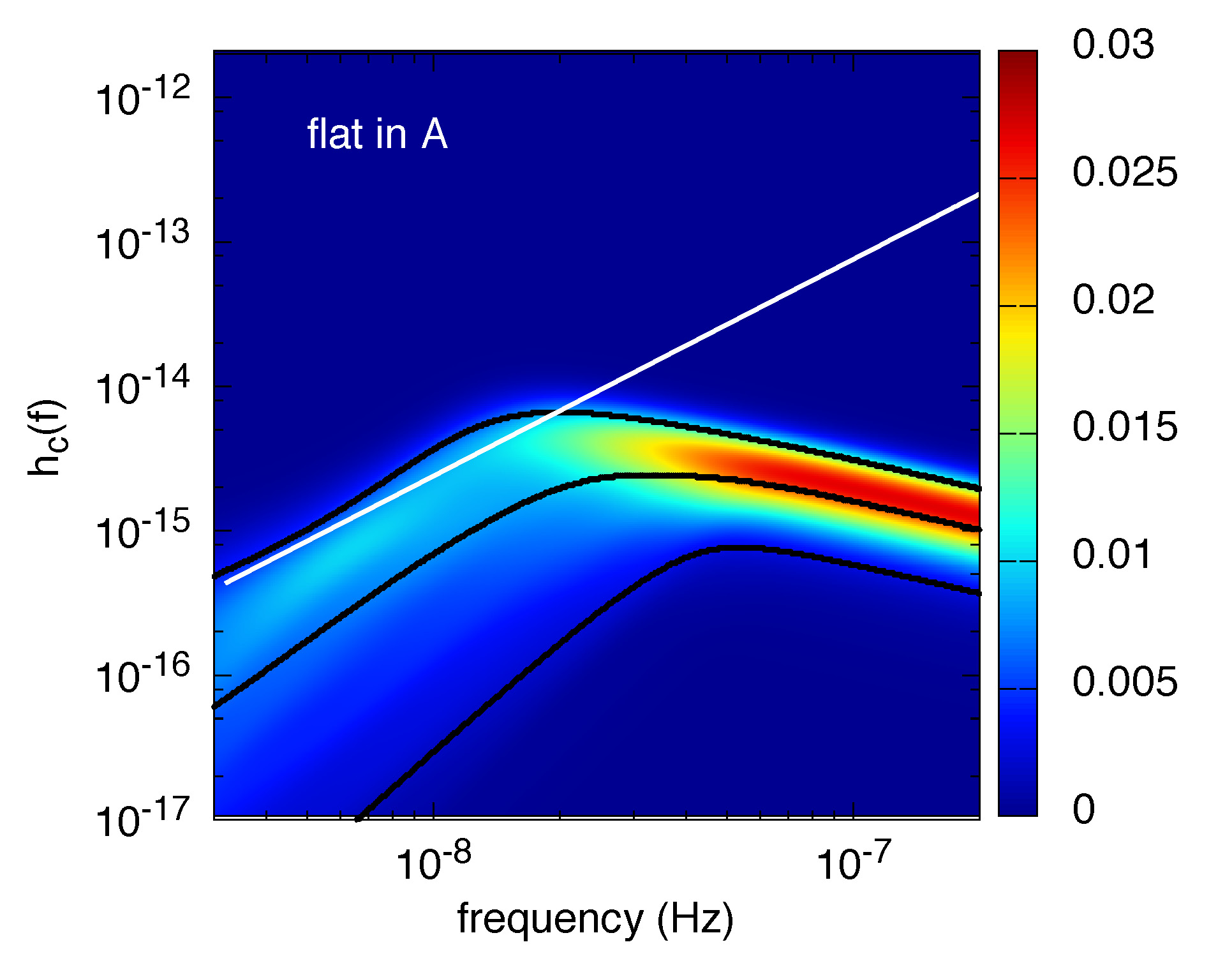} 
\includegraphics[clip=true,angle=0,width=0.48\textwidth]{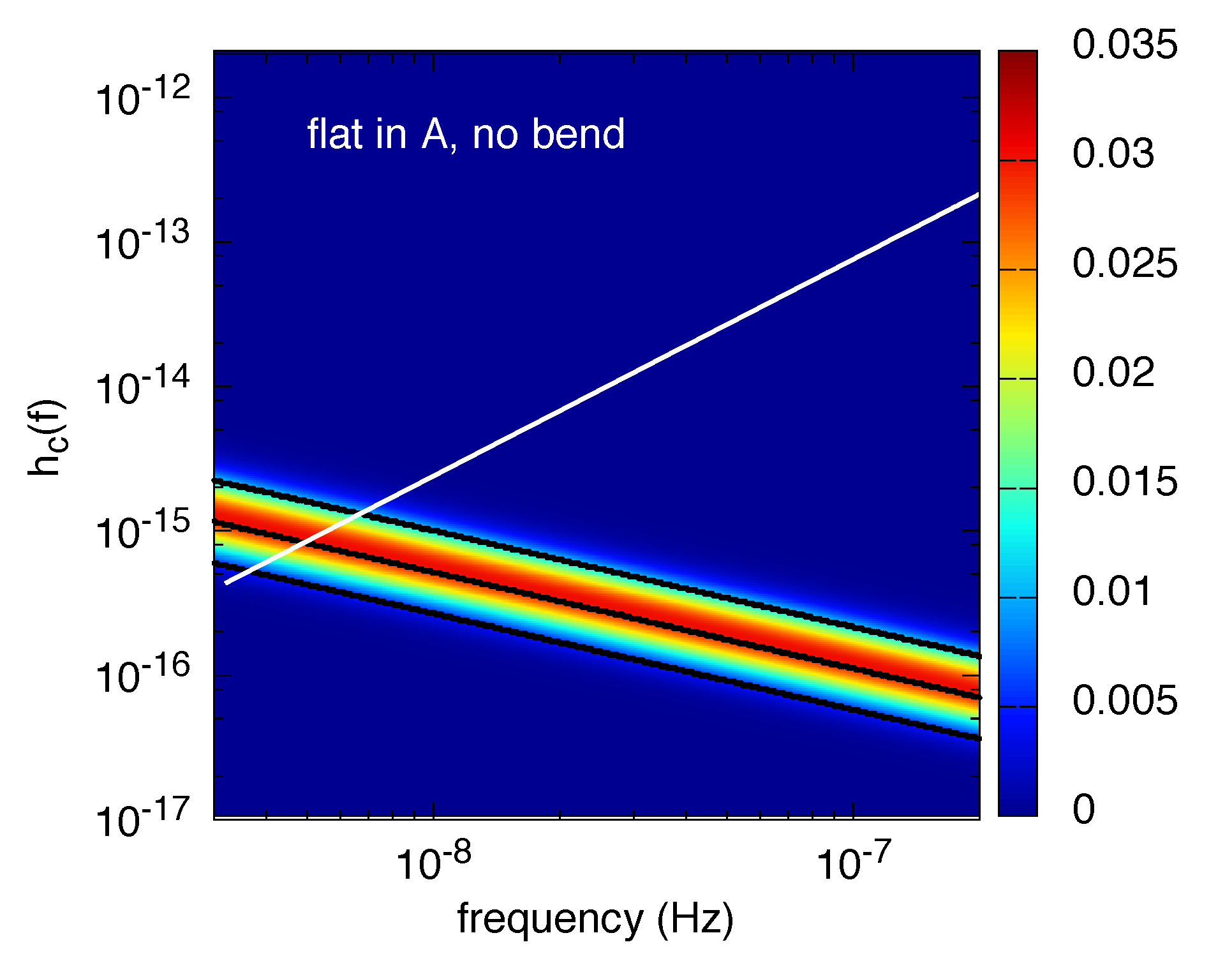} 
\caption{\label{fig:Bgramnoise} Density plots of the recovered spectra from MCMC runs including the full GW signal model (with $A, f_b$, and $\kappa$ all free), for three different choices of amplitude prior. The simulated data in this case contained only noise. The priors used are, clockwise from top left, the Model A prior, the Model B prior, and the prior uniform in $A$. Also shown are the simulated white noise (white line) and the $5\%$, $50\%$, and $95\%$ confidence regions for the recovered spectra.}
\end{figure*}
\begin{figure*}[htp]
\includegraphics[clip=true,angle=0,width=0.48\textwidth]{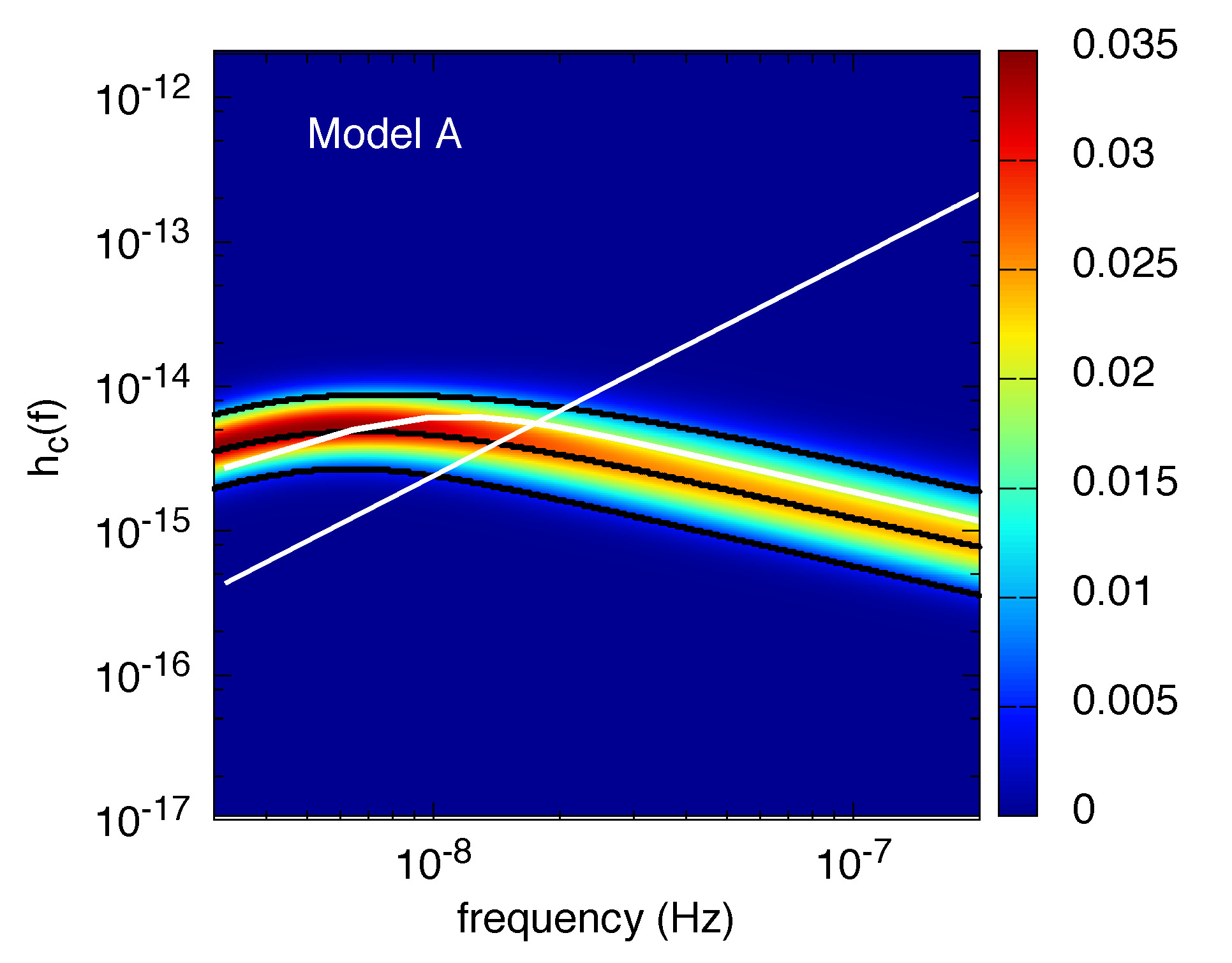} 
\includegraphics[clip=true,angle=0,width=0.48\textwidth]{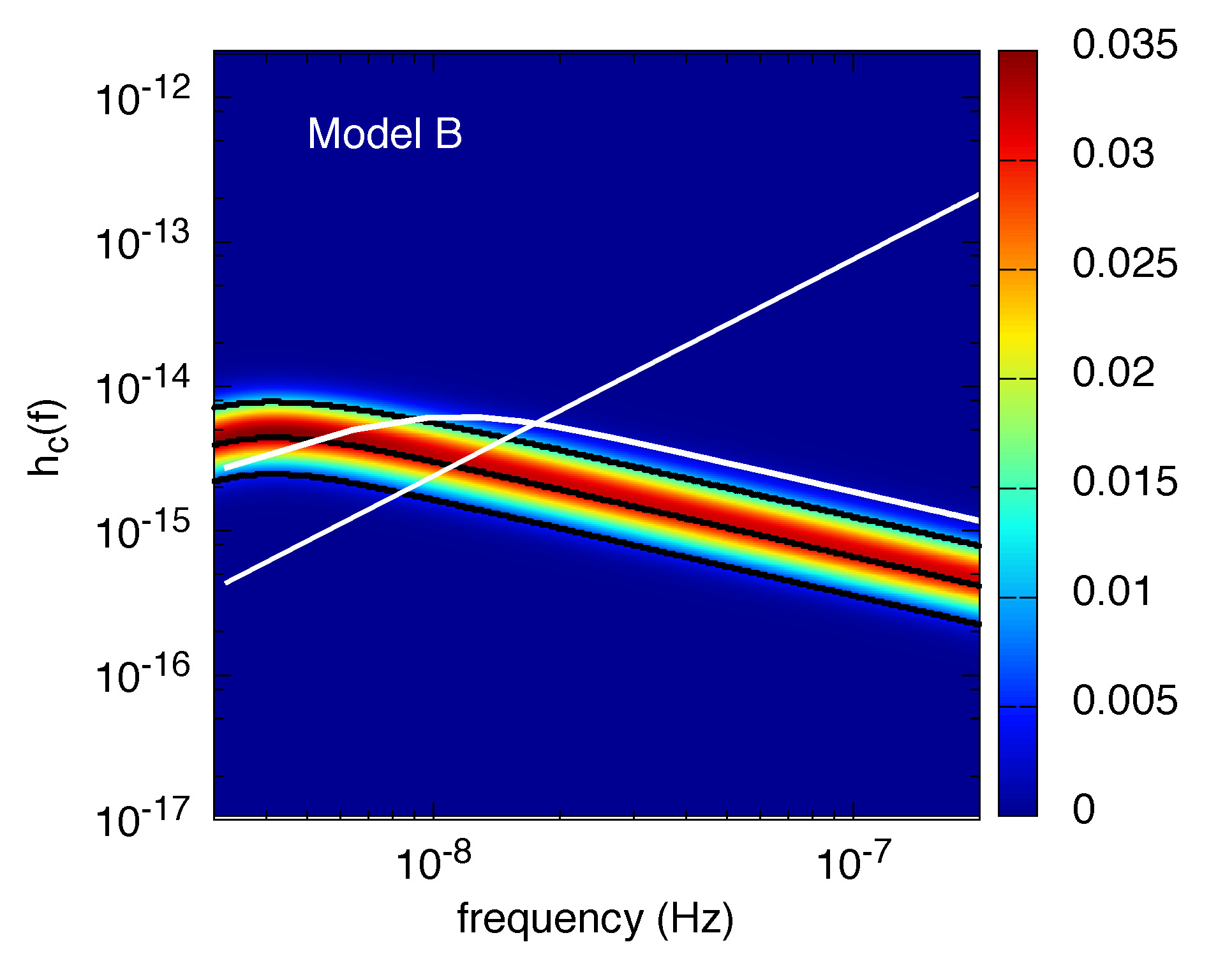} 
\includegraphics[clip=true,angle=0,width=0.48\textwidth]{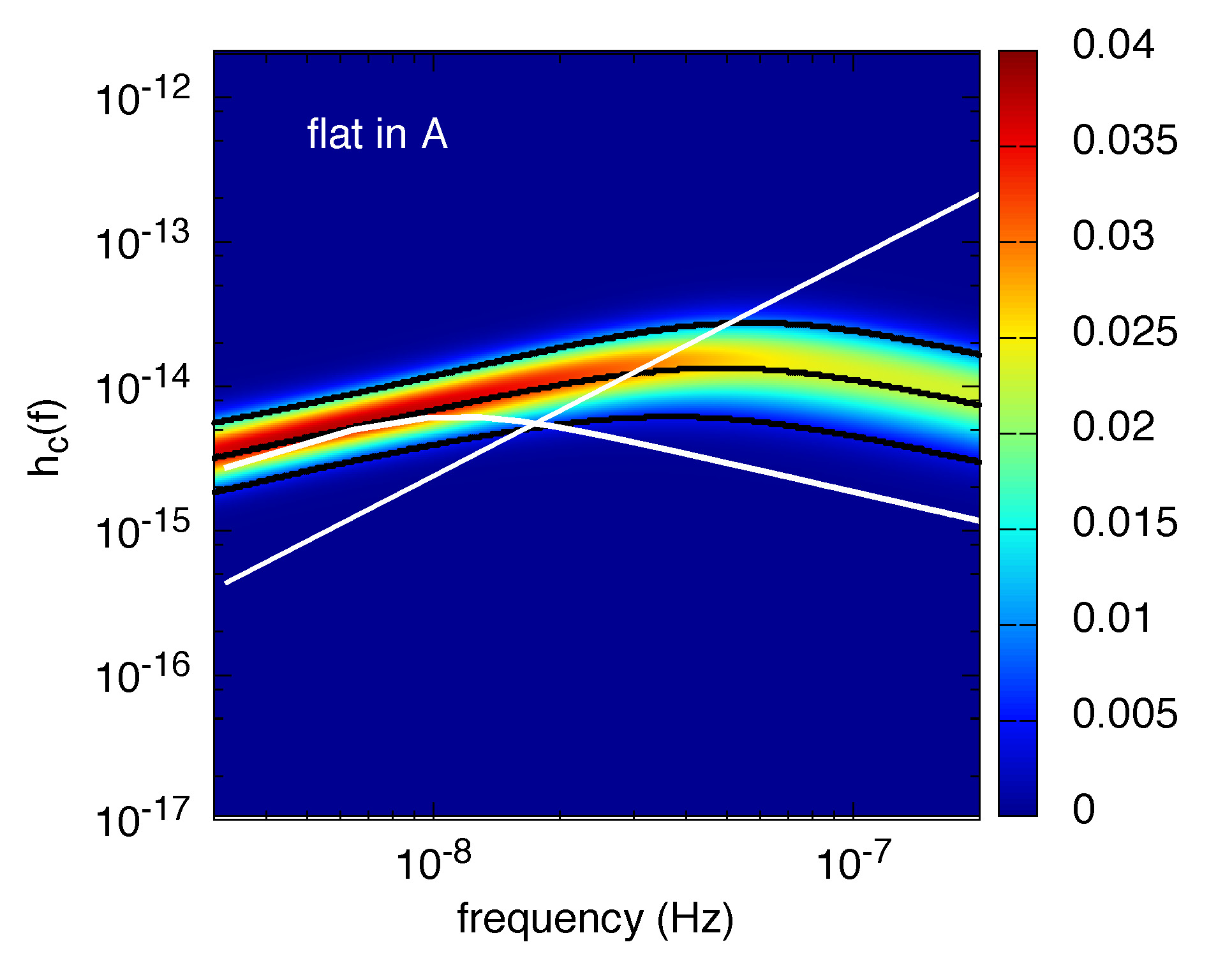} 
\includegraphics[clip=true,angle=0,width=0.48\textwidth]{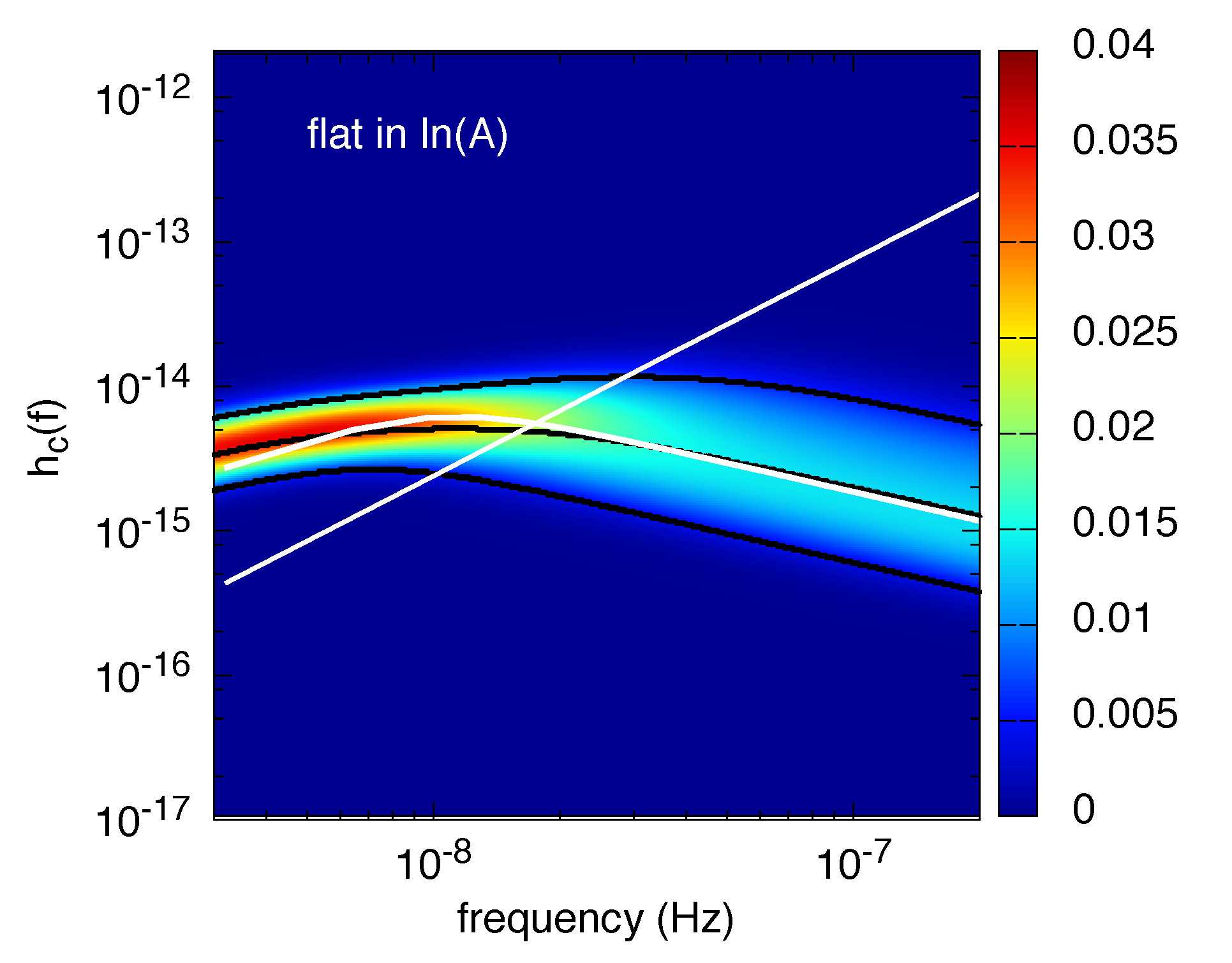} 
\caption{\label{fig:Bgraminj} Density plots of the recovered spectra from MCMC runs including the full GW signal model (with $A, f_b$, and $\kappa$ all free), for four different choices of amplitude prior. The simulated data in this case contained a GW background of type IV.c. The priors used are, clockwise from top left, the Model A prior, the Model B prior, a prior uniform in $\ln A$, and a prior uniform in $A$. Also shown are the simulated white noise (white line), the simulated signal (gray line), and the $5\%$, $50\%$, and $95\%$ confidence regions for the recovered spectra. The Model A prior and the prior uniform in $\ln A$ result in the best fir to the simulated spectrum.}
\end{figure*}

\begin{figure*}[htp]
\includegraphics[clip=true,angle=0]{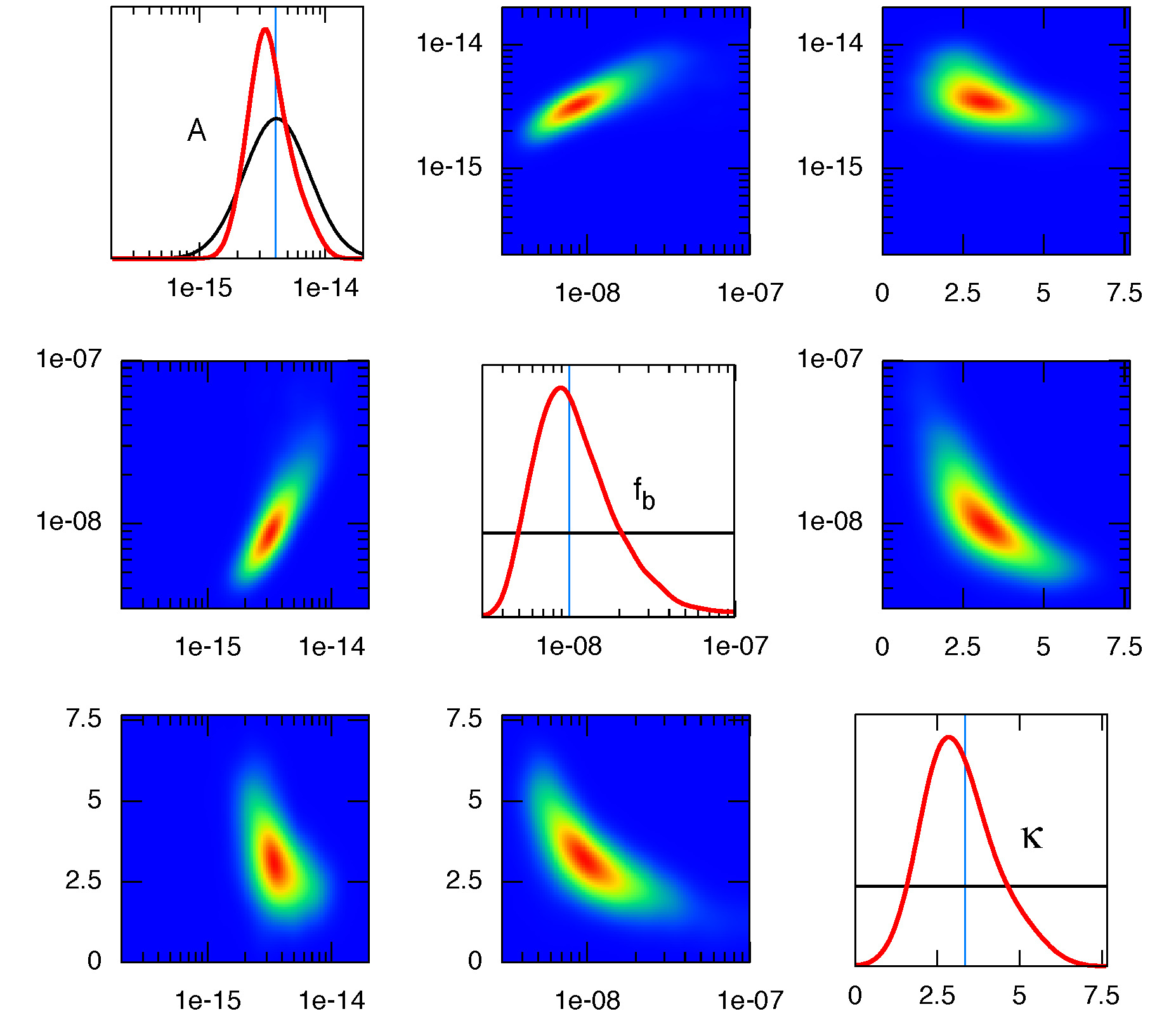} 
\hskip 0.5cm
\includegraphics[clip=true,angle=0]{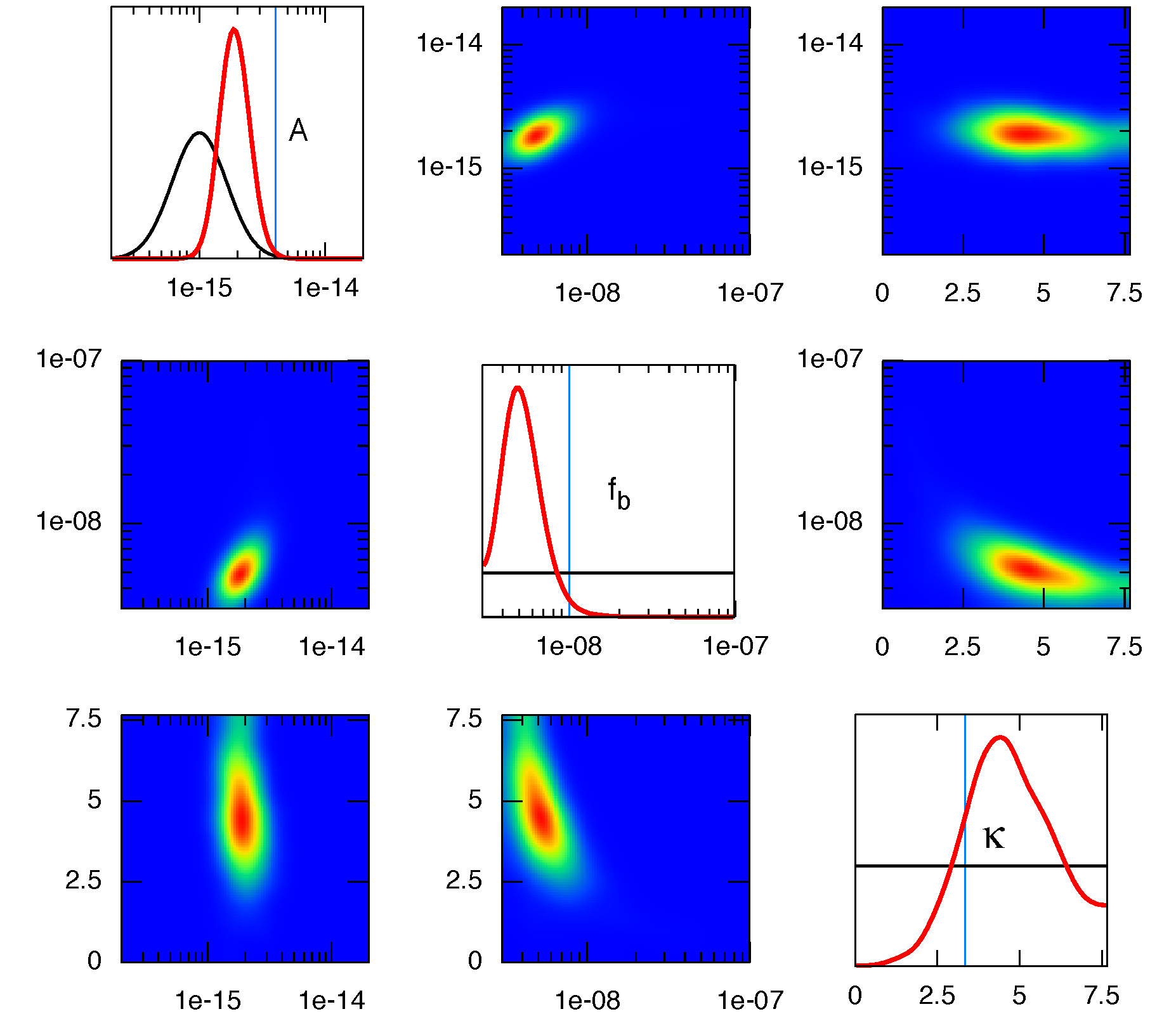} 
\vskip 0.5cm
\includegraphics[clip=true,angle=0]{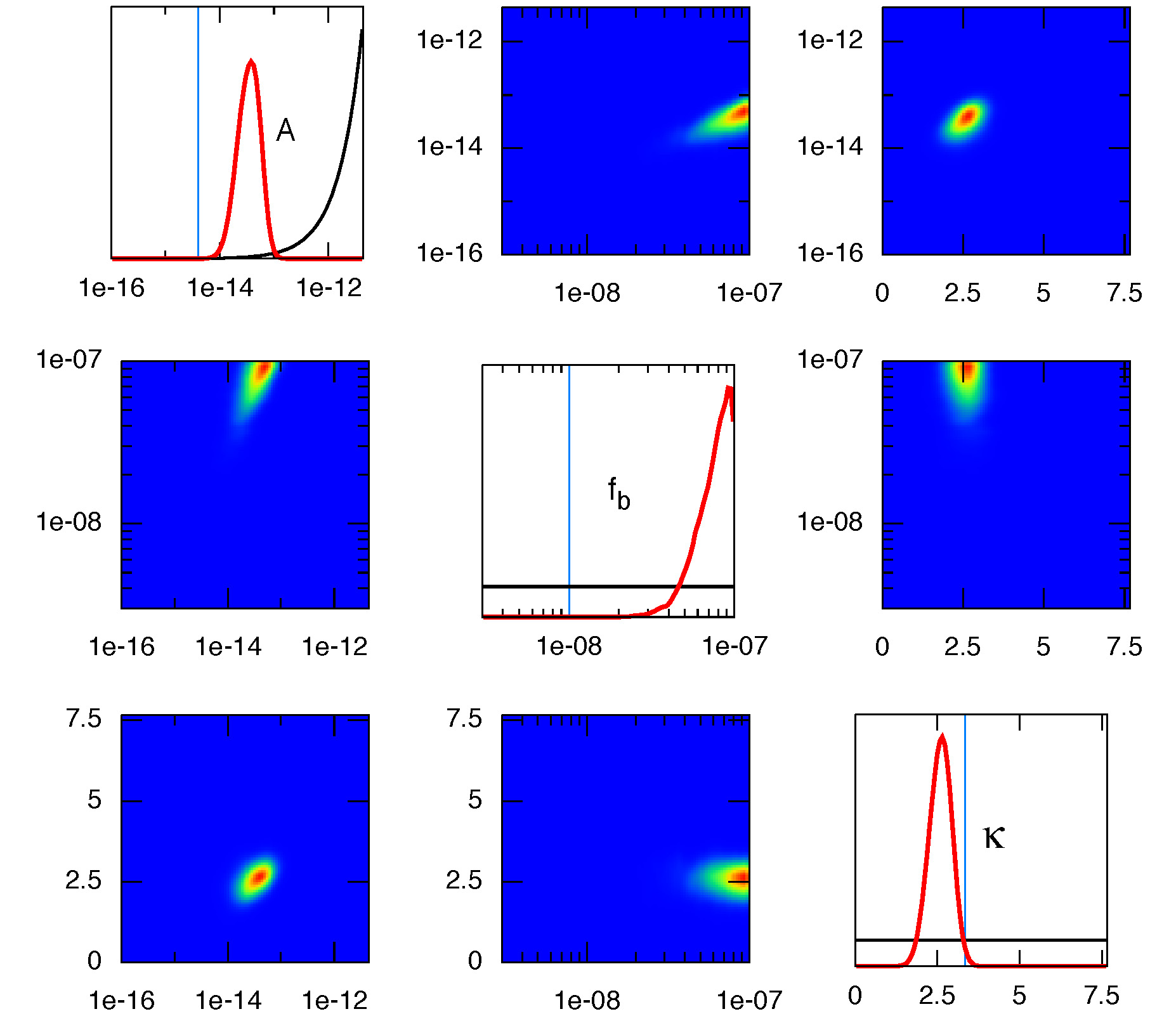} 
\hskip 0.5cm
\includegraphics[clip=true,angle=0]{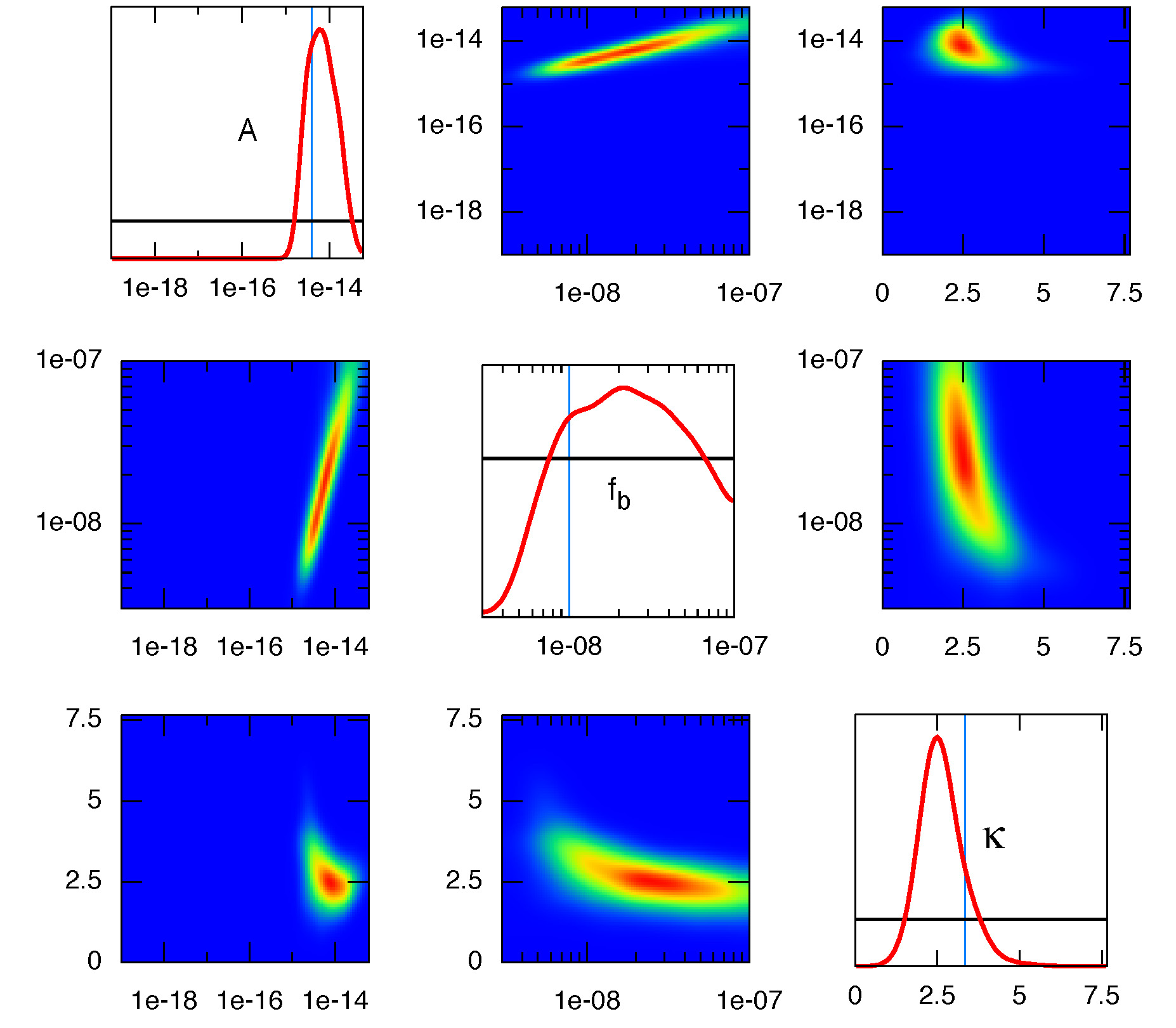} 
\caption{\label{fig:postpriorinjU} Prior and posterior distributions for the three GW parameters, $A, f_b$, and $\kappa$, recovered using four different amplitude priors. Along the diagonal are plotted the one-dimensional prior and posterior distributions for (from top) $A$, $f_b$, and $\kappa$. The off-diagonals show the 2-d posteriors. The priors used are, clockwise from top left: Model A, Model B, uniform in $\ln A$, and uniform in $A$. The simulated signal had $A=4.0\times10^{-15}$, $f_b = 10^{-8}$ Hz, and $\kappa = 10/3$. It is clear the prior that is uniform in $A$ leads to strong biases in the recovered parameter values. Thus, although this prior is a good choice for producing conservative upper limits, it has limited utility in parameter estimation studies.}
\end{figure*}

In addition to selecting between models, we are also interested in determining the accuracy with which the GW spectrum can be recovered.
As a first method for visualizing our ability to constrain the shape of the GW background, in Figure~\ref{fig:Bgramnoise} we show density plots of the recovered spectra calculated by simulating a noise-only signal, and recovering with the full GW spectral model. This is the type of analysis that leads to an upper limit on the GW amplitude, as there is no detection. To generate this plot, we calculate the spectrum at a sample of points in the Markov chain, then make a histogram of this spectrum at every frequency. The color map shows the weight of the histogram in each bin, with red corresponding to higher weight, and blue corresponding to lower weight. Also shown is the white noise level (as a white line), and the $5\%$, $50\%$, and $95\%$ confidence regions for the shape of the spectra. The purely GW-driven signal model with a uniform amplitude prior, labeled ``flat in $A$, no turn'' in
Figure~\ref{fig:Bgramnoise}, yields a 95\% upper limit on the amplitude of $A_{95} < 7.3\times 10^{-16}$. This number can be directly compared to the published upper limits from real PTAs, since these analyses also
assume a purely GW-driven evolution. The current upper limits are: the Parkes Pulsar Timing Array - $A < 2.4\times 10^{-15}$; the North American Nanohertz Observatory for Gravitational Waves (NANOGrav) - $A < 7\times 10^{-15}$; European Pulsar Timing Array - $A < 6\times 10^{-15}$. Thus, our simulated array is roughly three times more sensitive than the reported state-of-the-art, but based on projections~\cite{2013CQGra..30v4015S},
we expect our simulated sensitivity to be reached in reality within the next two to three years. The upper limits
on the amplitude from the models in Figure~\ref{fig:Bgramnoise} that allow for a bend in the spectrum are, unsurprisingly, significantly weaker. For example, the model with a constant amplitude prior and a bend in the
spectrum leads to an upper bound of $A_{95} <  6.8\times 10^{-15}$ - a full order of magnitude worse that the limit for the purely GW driven model.

Figure~\ref{fig:Bgraminj} shows posterior distributions for the recovered power spectra, but this time for simulated data includes an astrophysically-driven GW background. The parameters for this background were $A = 4.0 \times 10^{-15}$, $f_b = 3\times 10^{-8}$ Hz, and $\kappa = 10/3$. In this case, the simulated signal is plotted along with the white noise. It is again clear that the different choices in amplitude prior have a strong effect on the recovered spectrum. In this case, we have also included an example with a prior that is uniform in the logarithm of the amplitude. This choice of prior has been shown~\cite{Taylor:2014iua,2013CQGra..30v4004E} to be a poor choice for setting upper limits, but we suggest that it is a superior choice for parameter estimation. 

Note that the recovered spectra in Figure~\ref{fig:Bgraminj} do not fall precisely along the simulated track. This is an effect of the marginalization that occurs when calculating the densities. We are not showing the shape of any particular spectrum that was recovered by the chain. Rather, we are showing the distribution of the spectrum at each frequency, which is not required to have the same shape as the underlying spectral model. It is, essentially, a projection of the joint posterior distribution of the signal parameters $A$, $f_b$, and $\kappa$. We do not, therefore, expect the median recovered spectrum to lie directly on the simulated spectrum.

We will now discuss in detail the recovery of $A$, $f_b$, and $\kappa$. To examine our ability to measure the values of  these parameters, Figure~\ref{fig:postpriorinjU} shows the prior and posterior distributions for the signal parameters, for the same simulated signal as was used to generate Figure~\ref{fig:Bgraminj}, and for four choices of amplitude prior (uniform in $A$, uniform in $\ln A$, and the two astrophysical priors, Models A, B). Along the diagonal, we show the one-dimensional prior and posterior distributions for each parameter. In the off-diagonals, we show two-dimensional posterior distributions, illustrating the correlations between the parameters. The biases that arise from a uniform amplitude prior are clear in this figure - in particular, the amplitude, $A$, is pulled to very high values by this choice in prior. While this can be good for setting conservative upper limits on $A$ given the lack of a detection, it is clearly a poor choice for parameter estimation. In contrast, the prior uniform in $\ln A$, while bad for setting upper limits, is shown here to be a good choice for parameter estimation.

The maximum aposterior (MAP) parameter values, as well as the $90\%$ credible intervals, for $A,f_b$, and $\kappa$ are shown in Table~\ref{table:values} for all four choices of amplitude prior. From this table we see that only the Model A prior and the flat in $\ln A$ prior return credible intervals that contain the simulated value for $A$. Our model selection studies show that it would be very clear given a signal of this type that the Model A prior is favored, and consequently, is the model that should be used for parameter estimation in this case.

\begin{table*}
\centering
\begin{tabular}{ c  c  c  c  c  c c c c c}
\hline
      prior    & $A_{5\%}$ & $A_{\rm{MAP}}$ & $A_{95\%}$ & $(f_{b})_{5\%}$ & $(f_{b})_{\rm{MAP}}$ (Hz) & $(f_{b})_{95\%}$ & $\kappa_{5\%}$ & $\kappa_{\rm{MAP}}$ & $\kappa_{95\%}$ \\         
  \hline  
  Model A    & $2.05\times10^{-15}$ & $3.4\times10^{-15}$ & $7.10\times10^{-15}$ & $5.16\times10^{-9}$ & $9.4\times10^{-9}$ & $3.29\times10^{-8}$ & $1.65$ & $3.3$ & $5.17$ \\   
  Model B   & $1.35\times10^{-15}$ & $1.9\times10^{-15}$ & $2.83\times10^{-15}$ & $3.64\times10^{-9}$ & $5.0\times10^{-9}$ & $8.13\times10^{-9}$ & $2.6$ & $5.0$ & $7.01$ \\   
  flat in $A$    & $1.61\times10^{-14}$ & $5.2\times10^{-14}$ & $6.45\times10^{-14}$ & $4.53\times10^{-8}$ & $9.5\times10^{-8}$ & $9.83\times10^{-8}$ & $2.11$ & $2.6$ & $3.03$ \\   
  flat in $\ln A$    & $2.04\times10^{-15}$ & $8.8\times10^{-15}$ & $2.56\times10^{-14}$ & $6.16\times10^{-9}$ & $2.8\times10^{-8}$ & $7.96\times10^{-8}$ & $1.80$ & $2.5$ & $3.80$ \\   
   \hline  
\end{tabular}
 \caption{The maximum \emph{a posteriori} values for $A$, $f_b$, and $\kappa$, along with the $90\%$ credible intervals for each parameter, corresponding to the simulated signals and inferred distributions shown in Figure~\ref{fig:postpriorinjU}. Recall that the simulated values of these parameters were $A=4.0\times10^{-15}$, $f_b = 1\times10^{-8}$ Hz, and $\kappa = 10/3$. Note that only the Model A and uniform in $\ln A$ cases include the simulated values within their $90\%$ confidence intervals.}
 \label{table:values}
 \end{table*}
 
\subsection{Astrophysical Inference}
The measurement of $\kappa$ is of particular interest, as it maps most easily to astrophysical predictions. There are two approaches we can take to use the recovered $\kappa$ to distinguish between astrophysical models. The first is illustrated in Figure~\ref{fig:postpriorinjU}; we can simply produce posteriors of $\kappa$ given the detection of a GW background, and assess whether the measured value is consistent with a particular model. This presents a challenge, however, as the value of $\kappa$ is not measured with high precision. That is, the posterior distributions of $\kappa$ for both astrophysical amplitude priors and for the prior uniform in $\ln A$ have significant weight throughout the prior range. 

Additionally, the shape and peak of these posteriors depend strongly upon the choice of prior on $A$. This can, however, be viewed more as a feature than a drawback, as the choice of prior on $A$ encodes astrophysical information. This implies that the correct method for drawing astrophysical inferences from the measurement of the stochastic background is through model selection, where the models are defined both by $\kappa$ and by the prior on $A$.

\begin{figure}
\includegraphics[clip=true,angle=0,width=0.48\textwidth]{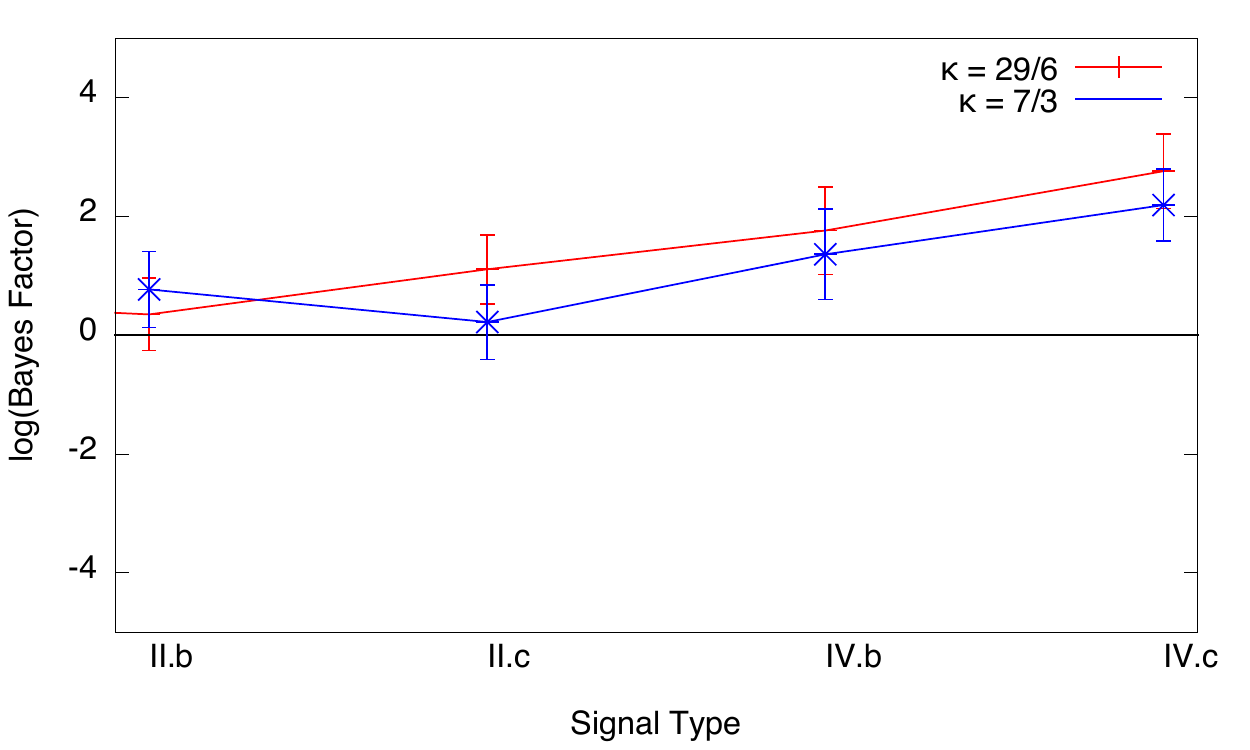} 
\caption{\label{fig:BFslope} The Bayes factors between a GW model with free parameters $A$ and $f_b$ and fixed $\kappa = 10/3$, and models with fixed $\kappa = 29/6$ (red) and  $\kappa = 7/3$ (blue) . The simulated signals (described in Table~(\ref{table:models})) all had $\kappa = 10/3$. The amplitude prior used here was the Model A prior. The signals are arranged along the x-axis in order of ascending detectability. A Bayes factor larger than unity indicates a preference for the $\kappa = 10/3$ model. For the most detectable signals, there is a clear preference for the correct model.}
\end{figure}

To investigate this type of model selection study, we simulated data of types II and IV, and recovered these simulated data sets using templates with $f_b$ free, but with fixed values for $\kappa$, using the amplitude prior for Model A. We then calculated the Bayes factors between these two models. The spectral models used for recovery have fixed slopes $\kappa = 10/3$, $\kappa=7/3$ or $\kappa = 29/6$. The
simulated data had a spectrum with $\kappa = 10/3$, which corresponds to stellar slingshot hardening. The $\kappa=7/3$ model corresponds to hardening due to a thin, cold circumbinary gas disk, while the
$\kappa = 29/6$ was randomly selected to provide a model with a shallower spectrum than was used to generate the data. Figure~\ref{fig:BFslope} shows the log Bayes factor of the $\kappa=10/3$ model
relative to the $\kappa = 29/6$ (red) and  $\kappa = 7/3$ (models).  So long as the bend frequency is high enough (signal type IV), and the amplitude large enough (sub-types b,c), it is possible to distinguish
between the different hardening mechanisms. 

\subsection{Learning from a Non-detection}

\begin{figure*}[htp]
\includegraphics[clip=true,angle=0]{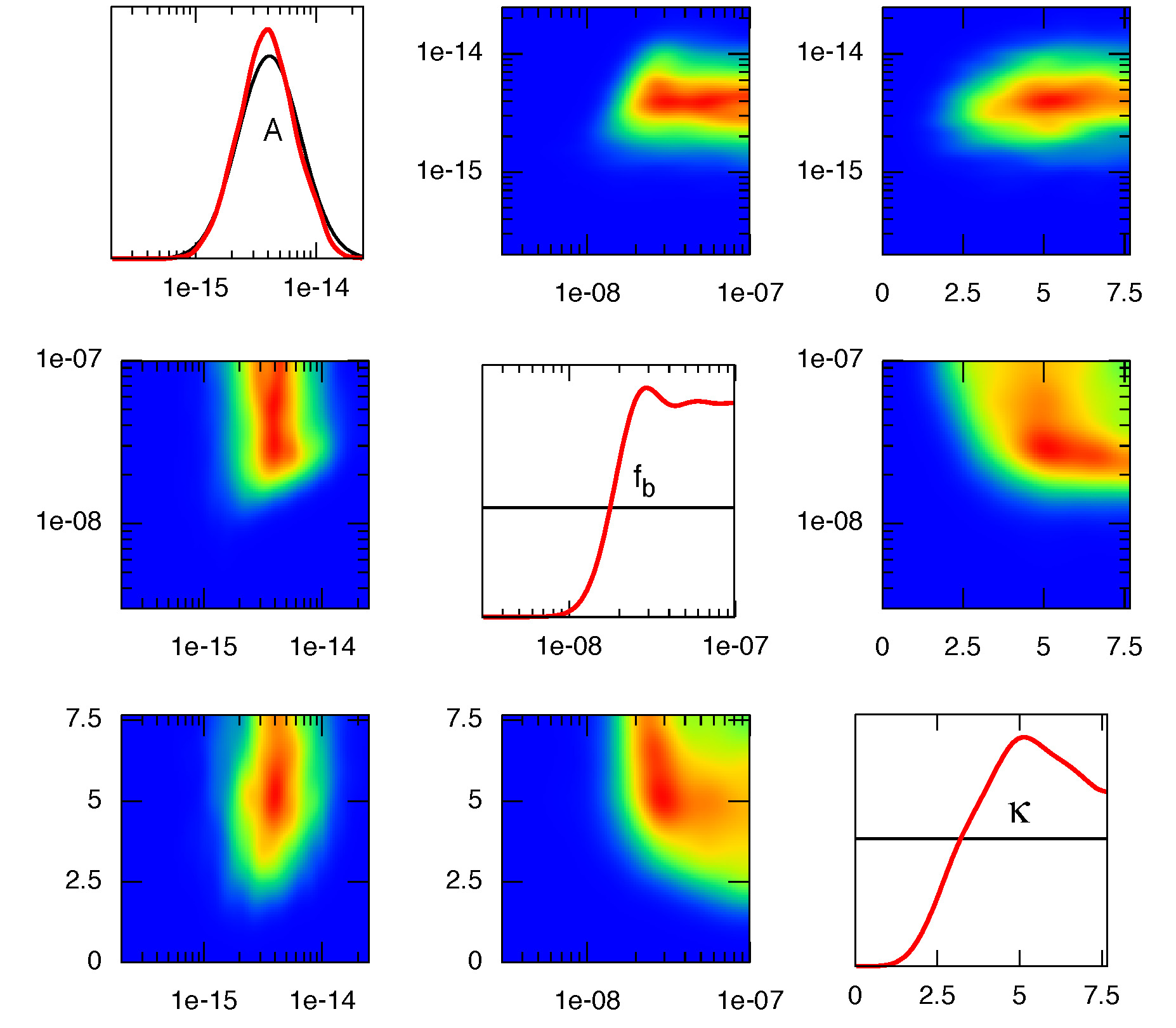} 
\hskip 0.5cm
\includegraphics[clip=true,angle=0]{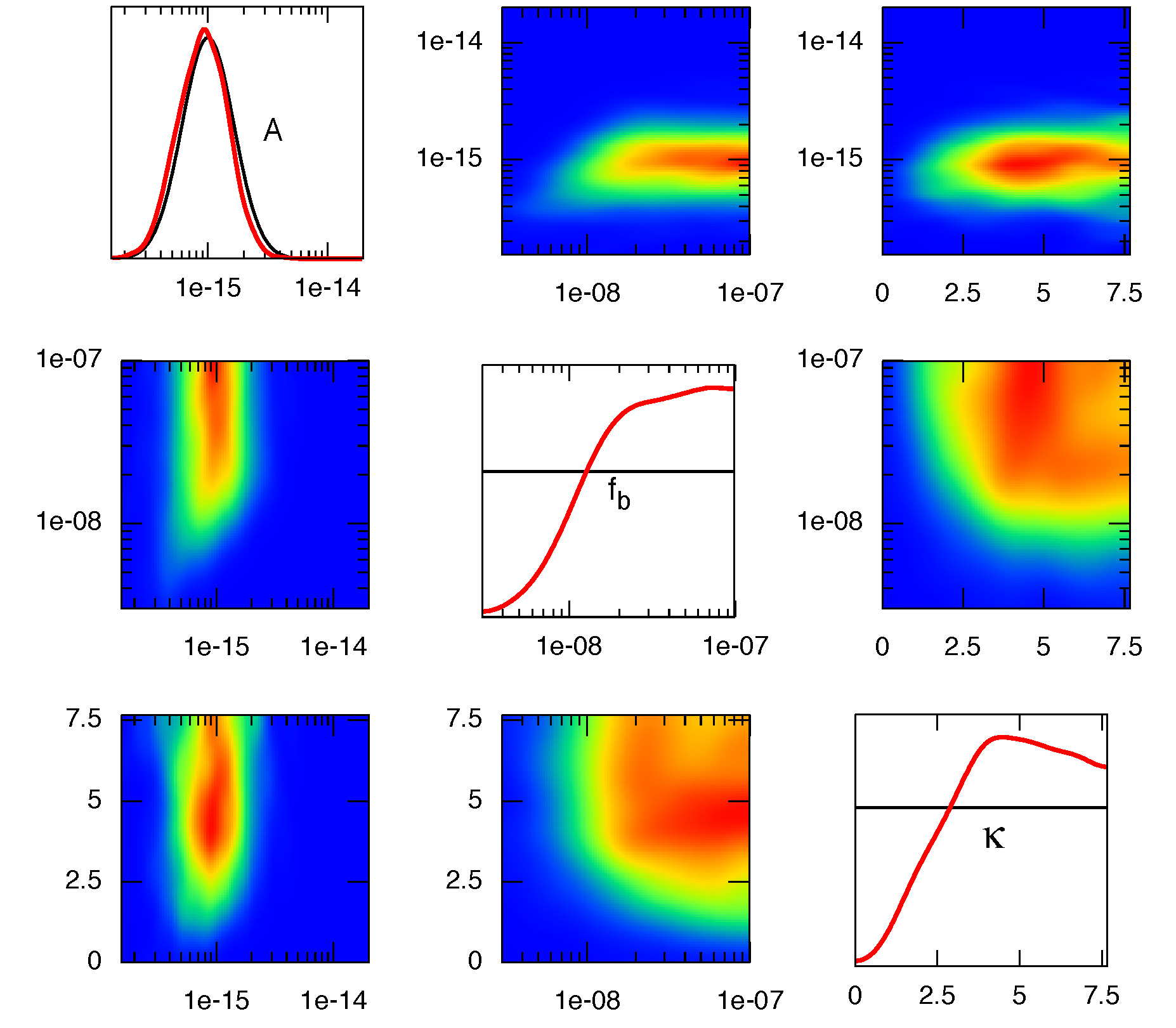} 
\vskip 0.5cm
\includegraphics[clip=true,angle=0]{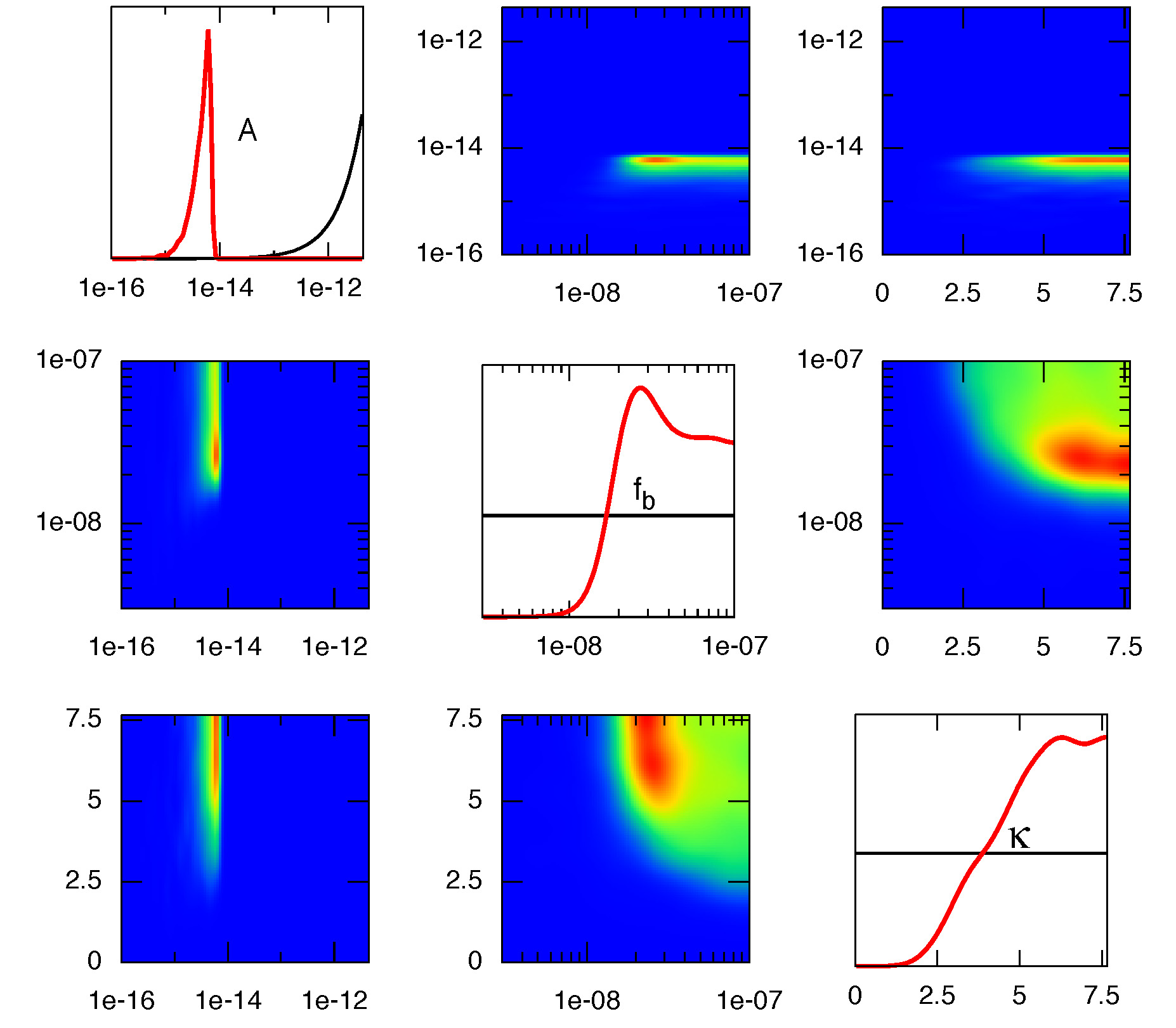} 
\caption{\label{fig:postpriorinjnoise} Prior and posterior distributions for the three GW parameters, $A, f_b$, and $\kappa$, recovered by analyzing simulated data that contains only white noise.  Along the diagonal are plotted the one-dimensional prior and posterior distributions for (from top) $A$, $f_b$, and $\kappa$. The off-diagonals show the 2-d posteriors for all combinations of these three parameters. The amplitude priors are, clockwise from top left: Model A, Model B, and uniform in $A$. Note that, even in the absence of a GW signal, the posterior distributions are substantially different from the priors. This indicates that we can learn about astrophysical models even if no detection is made.}
\end{figure*}

Information is gained from any measurement that leads to a posterior distribution that is different from the prior distribution.  In Figure~\ref{fig:postpriorinjnoise}, we show the prior and posterior distributions of $A, f_b$, and $\kappa$, as above, but now recovered from running on a signal that contains only noise. These distributions are shown for three different choices of amplitude prior - Model A, Model B, and uniform in $A$. It is clear that these distributions are significantly altered from the priors, showing that we can learn about astrophysical model space even lacking a GW detection. 

The primary result from PTA searches for a stochastic GW background is typically quoted as an upper limit on the amplitude $A$. Unsurprisingly, for the signal analyzed in Figure~\ref{fig:postpriorinjnoise}, the $95\%$ upper bound on $A$ depends on the choice of prior. For Model A, $A_{95\%} = 9.95\times10^{-15}$; for Model B, $A_{95\%} = 1.99\times10^{-15}$; and finally, for a uniform prior in $A$, $A_{95\%} = 6.83\times10^{-15}$. We see, then, that Model A leads to the most conservative upper bounds. Note that these bounds are for spectral models that allow for a bend in the spectrum, the limits assuming purely GW driven evolution are almost an order of magnitude lower. We will now explore precisely how much information we gain by a non-detection for the two astrophysical models.

To quantify exactly how much we can learn in this case, we compute the information gain (in bits) as the Kullback-Leibler (KL) divergence~\cite{RePEc:eee:csdana:v:54:y:2010:i:7:p:1719-1731,2013arXiv1308.6753V}
between the posterior and prior distributions. If the posterior matches the prior, we have learned nothing from the data. The larger the difference, the more we have learned.
The KL divergence between the posterior, $p(\vec{x}|d)$, and the prior, $p(\vec{x})$, is given by
\be
{\rm KL}(p(\vec{x}|d)||p(\vec{x})) = \int p(\vec{x}|d) \log \left( \frac{p(\vec{x}|d)}{p(\vec{x})}\right) d\vec{x}.
\label{Eq:KL}
\ee
This quantity can be calculated via thermodynamic integration, and is equal to
\be
{\rm KL}(p(\vec{x}|d)||p(\vec{x})) = E_{\beta=1} [\ln p(d|\vec{x})] - \ln p(d),
\ee
which is the expectation value of the log likelihood minus the log evidence. For the case illustrated in Figure~\ref{fig:postpriorinjnoise}, the KL divergences between posterior and prior are
${\rm KL}_{\rm Model \; A}= 308.1 \pm 0.4$ bits and ${\rm KL}_{\rm Model \; B}= 304.6 \pm 0.4$ bits. These numbers encode the information we have learned
about both the signal and the \emph{noise} parameters. One way to get a rough estimate of what we learned about the signal model is to compare the information gain for the signal and noise model to the information
gain for the noise model alone, which comes out to ${\rm KL}_{\rm Noise \; Model}= 301.2 \pm 0.4$ bits, suggesting an information gain of  $\sim 7$ bits for Model A and $\sim 3$ bits for Model B.
To properly isolate the information gained about the astrophysical models, we need to perform the integral in Eq.~(\ref{Eq:KL}) over only the prior and posterior of the signal parameters $A,\, f_b$, and $\kappa$. The functional form of the prior distributions is known, so this presents no difficulty. For the posterior distributions, we perform a three-dimensional KDE smoothing of the posterior distributions derived from our Markov chains. The KDE is applied
to chain samples that are mirrored at the prior boundaries to reduce edge effects. We then use these smoothed distributions to numerically integrate Eq.~(\ref{Eq:KL}). We find that, for Model A the information gain for the spectral model is ${\rm KL}= 1.5 \pm 0.08$ bits, while for Model B the information gain is ${\rm KL}= 0.7 \pm 0.04$ bits.
These numbers are significantly smaller than the crude estimate obtained by taking the difference between the noise and signal models, but this is not surprising - we know that there are significant correlations between the
noise parameters and the signal parameters, and the information measure is not additive. While these information gains are not large, they show that the posterior and prior distributions are measurably different, and
that we begin to learn about the astrophysical models driving the SMBH mergers even before a detection is made. As a point of comparison, the information gain for Model A is equal to the information gained about
cosmological models in going from the 7 year Wilkinson Microwave Anisotropy Probe (WMAP) data set to the 9 year data set, but is significantly less than the 30 bits gained in going from the
9 year WMAP maps to the higher resolution Planck maps~\cite{Seehars:2014ora}.

As the signal-to-noise ratio grows, the information gain grows. For the strong signal examples shown in Figure \ref{fig:postpriorinjU}, the information gains are ${\rm KL}= 3.1 \pm 0.2$ bits for Model A
and ${\rm KL}= 4.8 \pm 0.2$ bits for Model B. In this instance
we learn more about Model B since the amplitude of the simulated signal is large compared to what is predicted by the model, so there is a greater difference between the prior and posterior distributions.

An alternative way of seeing that we learn something about the astrophysical models from a non-detection is to compare the evidence for models that include a bend in the spectrum to those that
assume a purely GW driven evolution. For the noise-only data sets used to generate Figure~\ref{fig:postpriorinjnoise}, the log Bayes factor in favor of there being a bend in the spectrum is
$\ln {\rm BF} = 18.7 \pm 0.5$ for the Model A amplitude prior and $\ln {\rm BF} = 3.5 \pm 0.5$ for the Model B amplitude prior. These results say that a non-detection of GWs by a PTA with the sensitivity of our simulated array
would rule out purely GW driven evolution of the these merger models. Scaling back the sensitivity of the simulated array by a factor of two, (increasing the timing noise from 200 ns to 400 ns) to get something closer to
the NANOGrav sensitivity in 2015, yields $\ln {\rm BF} = 5.6 \pm 0.5$ for the Model A amplitude prior and $\ln {\rm BF} = 1.4 \pm 0.4$ for the Model B amplitude prior. At this lower sensitivity there would still be
strong evidence for non-GW driven evolution for Model A, but not for Model B.

\section{Summary}
\label{Sec:summary}
Pulsar Timing Arrays are likely to detect a stochastic GW background from binary SMBHBs before the end of the decade~\cite{2013CQGra..30v4015S}. The astrophysical processes that drive
SMBH binaries toward merger are not fully understood, and
are largely unconstrained by observations. Processes such as stellar scattering can drive the binaries through the sensitive band of PTAs more quickly than GWs, leading to a diminution of the GW signal. This may make the detection of GWs more challenging, but also opens up a new avenue for learning about the astrophysics of SMBH mergers through measurement of the spectral shape.

We have shown that a simple model for the spectrum, described in Eqs.~(\ref{Eq:pspec}), (\ref{hc}), are useful for the detection of GW backgrounds that are generated by SMBH binaries which are driven
by more than one type of mechanism within the pulsar timing band. This model can also be used in parameter estimation studies to characterize this GW background. 

We found that the choice of prior on the amplitude can significantly impact parameter estimation, and that the commonly used uniform prior in amplitude leads to especially large biases.
A prior that is uniform in the logarithm of the amplitude was found to be a far better choice.

We have shown that pulsar timing observations can be used to distinguish between models that are characterized by different priors on the amplitude of the GW background. In future work, we plan to
extend this study to more detailed models that make predictions about the slope parameters and the bend frequencies.

Finally, we have illustrated that information is gained about astrophysical models, even when no detection is made.

\acknowledgments

We have benefited from informative discussions with Justin Ellis, Xavier Siemens and useful feedback from members of the NANOGrav collaboration. LS was supported by Nicol\'{a}s Yunes'  NSF CAREER Award PHY-1250636. NJC was supported by NSF Award PHY-1306702

\bibliography{master}
\end{document}